\documentclass{article}
\usepackage{graphicx}  
\usepackage{amsmath}   
\usepackage[compress]{cite}
\usepackage{amssymb}   
\usepackage{bm} 
\usepackage{dcolumn}
\usepackage{color}
\usepackage{mathrsfs}
\usepackage{amsfonts}
\usepackage{varioref}
\usepackage[margin=1in]{geometry}
\RequirePackage[colorlinks,citecolor=blue,urlcolor=magenta,linkcolor=blue]{hyperref}
\usepackage{tikz}
\usepackage{lmodern}
\usetikzlibrary{decorations.pathmorphing}
\tikzset{snake it/.style={decorate, decoration=snake}}
\usepackage{fancybox}
\expandafter\ifx\csname package@font\endcsname\relax\else
\expandafter\expandafter
\expandafter\usepackage
\expandafter\expandafter
\expandafter{\csname package@font\endcsname}%
\fi
\DeclareFontFamily{OT1}{pzc}{}
\DeclareFontShape{OT1}{pzc}{m}{it}%
             {<-> s * [0.900] pzcmi7t}{}
\DeclareMathAlphabet{\mathscr}{OT1}{pzc}%
                                 {m}{it}
\labelformat{section}{Section #1} 
\labelformat{subsection}{Section #1} 
\labelformat{equation}{Eq.(#1)} 
\labelformat{figure}{Fig.~#1} 
\labelformat{subfigure}{Fig.~\thefigure#1} 
\labelformat{table}{Tab.~#1} 
\newcommand{\be}{\begin{equation}}
\newcommand{\ee}{\end{equation}}
\newcommand{\bea}{\begin{eqnarray}}
\newcommand{\eea}{\end{eqnarray}}

\def\I{\mathcal{I}}
\begin{document}
 
\title{Decoding infrared imprints of quantum origins of black holes}

\author{Sumanta Chakraborty\footnote{sumantac.physics@gmail.com}$~^{1}$ and Kinjalk Lochan\footnote{kinjalk.lochan@gmail.com}$~^{2}$
\\
$~^{1}${\small{School of Physical Sciences, Indian Association for the Cultivation of Science, Kolkata 700032, India}}\\
$~^{2}${\small{Department of Physical Sciences, IISER Mohali, Manauli 140306, India}}}

 
\maketitle
\begin{abstract}
We analyze the emission spectrum of a (fundamentally quantum) black hole in the Kerr-Newman family by assuming a discretization of black hole geometry and the holographic entropy-area relation. We demonstrate that, given the above structure of black hole entropy, a macroscopic black hole always has non-continuously separated mass states and therefore they descend down in discrete manner. We evaluate the step size of the discrete spectrum, which vanishes in the extremal limit, leading to a continuum spectrum as expected from thermal nature of black holes. This further reveals an interesting relation, in each class, between the dynamic and kinematic length scales for all black holes belonging to the Kerr-Newman family, pointing towards a possible universal character across the class, dependent only on black hole mass. Further, we have presented the computation of maximum number of emitted quanta from the black hole as well as an estimation of its lifetime. We also argue the independence of these features from the presence of additional spacetime dimensions.
\end{abstract}

\section{Introduction and Motivation}

Realization of the fact that black holes behave as thermodynamical objects \cite{Bekenstein:1974ax,Bardeen:1973gs,Gibbons:1977mu,Hawking:1976de, Bekenstein:1973ur,Bekenstein:1972tm,Hawking:1974sw,Jacobson:1995ab,Padmanabhan:2009vy,Padmanabhan:2013nxa} has, since long, been fuelling the hope of reconciling the quantum theory with the framework of gravity in a successful and consistent manner. In particular, the thermodynamic nature of black holes suggest a strong connection between macroscopic geometrical constructs and microscopic quantum behaviour. Thus demystifying and understanding the black hole geometry has been one of the pioneering goals of modern physics, which in all likelihood, will provide an useful insight to the quest of arriving at a successful quantum theory of gravity \cite{Maldacena:1996ky,Rovelli:1996dv,Dowker:2005tz,Ashtekar:1997yu}. Different approaches towards quantizing gravity \cite{Maldacena:1996ky,Rovelli:1996dv,Dowker:2005tz,Ashtekar:1997yu}, emerging out since the discovery of black hole thermodynamics, have been accordingly oriented, keeping the hope of unifying gravity and quantum theory in sight. Unfortunately, despite decades of efforts along numerous directions, which have resulted into many promising candidates for quantum gravity, we have not been able to come up with a theory that can actually provide a consistent description to unify quantum theory and gravity. Each of these candidates has offered its own different pathway of studying and understanding different constructions of gravity, more particularly the black holes. However with no black holes available to experiment with, and a plethora of quantum gravity theories with their own predictions, the black hole mysteries \cite{Hawking:1976ra,Visser:2014ypa,Mathur:1997wb,Mathur:2009hf,Hooft:2015jea,Chakraborty:2017pmn,Hawking:2016msc,Hawking:2016sgy,Modak:2014qja} have led this search to a cross road.

The detections of gravitational waves \cite{Abbott:2016blz,Abbott:2016nmj} from merger of two black holes have recently ignited the debates regarding the inner workings of black holes \cite{Hawking:1976ra,Visser:2014ypa,Mathur:1997wb,Mathur:2009hf,Hooft:2015jea,Chakraborty:2017pmn,Hawking:2016msc,Hawking:2016sgy,Modak:2014qja,Kraus:2015zda,Hawking:2014tga,Frolov:2014wja,Vaz:2014era} and its possible implications for quantum gravity theories. There have been studies \cite{Abedi,Foit:2016uxn} to explore the imprints of some non-intuitive quantum features of the black holes, which appear in many frameworks of quantum gravity.  Therefore, the dynamical evolution of black holes can really act as a lead for testing of our theoretical models claiming to quantize gravity \cite{Cardoso:2017cqb,Maselli:2017kic,Cardoso:2017cfl}.

Apart from the classical gravitational wave emissions, there is also a quantum channel by which a black hole can emit, known as the famous {\it Hawking radiation} \cite{Hawking:1974sw}. There is a renewed interest in the quantum emission profile of black holes after the realization gained ground that quantum emission of black holes may well be hiding its intrinsic quantum face \cite{Alonso-Serrano:2015trn, Lochan:2015oba,Lochan:2016nbs,Chakraborty:2016fye}. Studies of such quantum emissions hold their own ground as the tussle between non-compatibility of the quantum theory with principles of gravity become manifest here, thanks to the thermal nature of the radiation \cite{Chakraborty:2017pmn,Saini:2015dea,Chen:2014jwq}. 

In this work, we will assume that the notion of black hole entropy as perceived in the context of thermodynamics of macroscopic black holes holds true through the microcanonical counting, in the quantum domain (see, for example  \cite{Bekenstein:1995ju,Bekenstein1974,Bekenstein:1997bt}). Given the above assumption on the nature of black hole entropy at a microscopic level we would like to address the question, across these class of theories, is there any signature of the underlying quantum nature of black holes which gets fossilized in its emission? In previous works \cite{Lochan:2015bha,Chakraborty:2016fye}, we have explicitly demonstrated that if black holes are indeed a macroscopic realization of a microscopic quantum geometry and the thermodynamic relation between entropy and area still holds in the microscopic level having holographic properties, then it is highly unlikely that the emission from them will have the suggested thermal behaviour. If the geometry (of black hole spacetime) gets discretized {\it in any fashion}, the emission lines from the macroscopic black hole become hugely discrete in the low frequency regime. Similar conclusions were reached in many other contexts and their far reaching implications are still being investigated \cite{Gray:2015pma,Hod:2015wva,Hod:2016xbu,Hod:2016rmg}. 

However the scenario depicted in earlier works was for a Schwarzschild black hole, with mass as the only hair \cite{Bekenstein:1995ju,Hod:1998vk,Hod:2000it,Barreira:1996dt,Lochan:2015bha,Hod:2015wva,Hod:2016xbu,Hod:2016rmg,Chakraborty:2016fye}, while astrophysical black holes are most likely to inherit non-zero angular momentum as well. Thus in the process of decay, the black hole loses both mass and angular momentum and as we will show in this work, the inclusion of angular momentum modifies the quantum spectrum of a black hole drastically. In particular, as we will depict, there will be lot of interesting physics happening near the extremal limit, which was non-existent for Schwarzschild black hole such as occurrence of a dense emission spectrum. Further, since black holes are supposed to be thermodynamical objects in classical theory, the emission lines must be consistent with both thermodynamics and the underlying quantum structure. As we will see, these two demands are strong enough that result into a discrete quantum spectrum for black holes \cite{Bekenstein:1995ju,Hod:1998vk,Hod:2000it,Lochan:2015bha,Chakraborty:2016fye}. This fact has been aptly demonstrated in the so called \emph{Bekenstein-Mukhanov} effect, showing the existence of a minimum frequency and hence the largest wavelength that a black hole can emit. Note that the above result is based on the assumption that the entropy-area relation, obtained from the thermodynamics of macroscopic black holes, comes through in {\it exact form} from the underlying quantum counting of this macroscopic realization of the quantum discrete geometry. However, typically there can be sub-leading corrections to such a relation, which will come through the exact (say) von-Neumann counting. In \ref{AppB}, we have addressed a class of such subleading correction which appear across many theories of quantum gravity \cite{Lochan:2012in,Lochan:2012sp,Sen:2012dw,Ghosh:2004rq}. Interestingly, for this class of corrections too, the results obtained in this work remain valid. Once we have minimum frequency of emission, we can track the dynamics of the black hole evaporation, counting the {\it maximum possible} number of emissions as well as the time taken by the black hole to turn Planckian in size where the laws of emission are supposed to get compromised.

The paper is organized as follows: In \ref{GeometricOp} we analyze the relation of the geometric aspects of a Kerr-Newman black hole in order to determine the mass gap between the nearest macroscopic configuration emerging out from a quantum operator which is discretized in integer steps along with the universal relation connecting various length scales. Subsequently, in \ref{Lifetime}, we estimate the maximum number of quanta that a Kerr-Newman black hole can emit and hence will provide a rough estimate of the lifetime of a black hole. Some discussions regarding extremal limit in this discretization scheme has been presented in \ref{ExtremalLimit}. We finally argue the robustness of our results for higher dimensional rotating as well as charged black holes in \ref{HighDim}, before summarizing our main findings in \ref{Conclusion}. 
\section{Holography and quantum spectrum of black holes} \label{GeometricOp}

We will  work with a premise that a quantum theory of gravity discretizes geometry in any (hitherto unknown) fashion. There are many candidate theories with different geometric operators getting quantized as well as with different spectrum of quantization \cite{Rovelli:1994ge,Vaz:2007td,Vaz:2009jj}. The results obtained in this paper remain fairly insensitive to the details of the geometry quantization as they care only about the {\it course gained} quantities at the macroscopic levels (by which we mean fairly away from being Planck sized).  Throughout this work we will exclusively concentrate on the Kerr-Newman family of black holes, which is constructed at the macroscopic level from any of the geometric quantization theories,  characterized by the mass $M$, angular momentum $J$ and charge $Q$ with the horizon located at,
\begin{align}
r_{\rm h}=M\left[1+\sqrt{1-\left\{\left(\frac{a}{M}\right)^{2}+\left(\frac{Q}{M}\right)^{2}\right\}}\right]~,
\end{align}
where $a=J/M$ is the rotation parameter associated with the Kerr-Newman black hole. Since we are interested in black hole spacetime, in what follows we will assume that $(a^{2}+Q^{2})<M^{2}$. Note that in the limit $\{(a/M)^{2}+(Q/M)^{2}\}\rightarrow1$ the term inside square root vanishes and is known as the extremal limit. 

In this context of Kerr-Newman black hole we will assume that certain geometric feature, (e.g., area) gets discretized, without referring to any particular scheme of quantum gravity. In this approach we will consider any classical black hole to be a macroscopic realization of the underlying quantum theory, which still awaits discovery. Remarkably enough, our results will be independent on the nitty gritties of such a quantum theory, except for the fact that some geometrical object has been discretized. Let such a geometrical construct, denoted by $\mathcal{Z}$ is assumed to have a discrete spectral profile characterized by $f(j)$, where $j$ symbolizes the quantum level, which, motivated by the compactness of the horizon we consider to be discrete. The particular form for $f(j)$ will depend on the details of the discretization scheme, which we will not be bothered about and shall keep it arbitrary.

The classical geometry, which in the quantum theory of gravity will be given by some state $\hat{\rho}$ and is understood to emerge in the macroscopic limit, when the expectation value of the geometric quantity $\mathcal{Z}$ becomes 
\begin{align}\label{Arb_Spec}
{\cal Z}= \text{Tr}[\hat{\rho}\hat{\cal Z}] =\alpha \sum_jn_j f(j)~,
\end{align}
where the fluctuations are understood to be heavily suppressed for macroscopic size system.
As mentioned earlier the discretization spectrum $f(j)$ as of now will be arbitrary, which will be known only when the correct quantum theory of black holes becomes available. The parameter $\alpha$ is a new scale which is brought along by the underlying quantum theory and is possibly related to the Planck scale. Further, the presence of a minimal Planck length in any theory of quantum gravity tells us that ${\cal Z}$ must be bounded from below.

In the macroscopic limit let the area of the event horizon get related to this characterizing geometric parameter as 
\begin{align}\label{A-Z Rel}
A=\kappa{\cal Z}^{\gamma}=\kappa\alpha^{\gamma}\Big\{\sum_{j}n_{j} f(j)\Big\}^{\gamma}~, 
\end{align}
where $\gamma$ is assumed to be some real positive number\footnote{We can generalize the relation of the area to the discretized variable in a polynomial fashion. Appendix A discusses the generalization.}. For example, when $\mathcal{Z}$ corresponds to square root of black hole area, then $\kappa=1$ and $\gamma=2$. If, we further accept that a macroscopic description emerging from the basic quantum theory should respect the Bekenstein's entropy-area  relation, then there must also be a way to connect entropy with the quantum description, such that,
\begin{align}\label{Holography}
S=\frac{A}{4}=\frac{\bar{\alpha}^{\gamma}}{4}\Big\{\sum_jn_j f(j)\Big\}^{\gamma}
=\ln g(n)~,
\end{align}
where $\bar{\alpha}^{\gamma}=\kappa\alpha^{\gamma}$ and $g(n)$ is effective number of micro states giving rise to the macroscopic black hole configuration, as we consider the black hole as a system in a micro-canonical set up. Many quantum gravity theories attempt to visualize this number of available microstates through their counting schemes and have been successful in doing so \cite{Sen2014,Rovelli:1996dv,Kaul:2012pf}. Therefore, recovery of this relation from microscopic counting does not really serve as a very strong discriminator between the available quantum gravity models. However, we will see (in \ref{MassGap}) that characterizing the nearest allowed configuration may well be such a tool. Furthermore one can argue that this feature remains true in different ensemble pictures of analysis as well. From \ref{Holography} it is simple to compute the number of micro states, leading to,
\begin{align}\label{IntegerConstraint}
g(n)=\exp{S}=\exp\left({\frac{A}{4}}\right)=\exp\left[{\frac{\bar{\alpha}^{\gamma}}{4} F^{\gamma}}\right];\qquad F\equiv \Big\{\sum_jn_j f(j)\Big\}~.
\end{align}
Clearly, as the number of microstates in the micro-canonical picture has to be an integer, \ref{IntegerConstraint} demands $\mathcal{B}F^{\gamma}/4=\ln K$ for some variable integer $K$, corresponding to different area values of the horizon. This can happen in two possible ways --- (a)  $\mathcal{B}/4 =\ln \mathcal{I}_{0}$ and $F^{\gamma}=\I$; or --- (b) $\mathcal{B}/4 =\I_0$ for some fixed integer $\I_0$, and $F^{\gamma} =\ln \I$ for variable integer $\I$, for all possible $\{n_j\}$ (at least) for which $\sum_j n_j \gg 1$, i.e., in the macroscopic limit. The choice (a) essentially  presents a integer shift \cite{Bekenstein:1995ju,Bekenstein:1997bt} in the macroscopic expectation of the geometric variable, while choice (b)  corresponds to a logarithmically discretized geometric operator \cite{Visser:1992ck}. We will discuss the scenario depicted by case (a) alone, i.e., in which $A=\mathcal{B} \I$. Here $\mathcal{B}$ carries the imprint of the underlying quantum structure and the integer $\I$ marks the micro states $\{n_{j}\}$ in a collective manner. Following the footsteps, the case (b) can also be taken up with equal ease and the results remain qualitatively similar to case (a), which we would not report here. Such an integerly discretized scheme may be advocated on various grounds including quantum gravity as well \cite{Barreira:1996dt,Rovelli:1996dv,Rovelli:2017mzl,Ashtekar:1997yu,Chakraborty:2017s}. As the black hole makes a transition from one macroscopic configuration characterized by $\{n_{j}\}$ to another macroscopic state denoted by $\{n_{j}'\}$, the integer changes from $\I$ to $\I'$.

Before concluding this section we would like to emphasize that the quantization scheme presented here is completely general, since we have \emph{not} assumed any particular form for the function $F(j)$ determining the quantization levels of black holes. As and when a viable theory of quantum gravity becomes available, it will certainly produce some $F(j)$ and the results derived in this work being independent of $F(j)$, will be readily applicable. Hence all the results we will derive in this work will have applicability for any theory of quantum gravity, as long as it predicts quantization of certain geometrical quantity. It is \emph{not at all dependent} on the details of the quantization procedure.
\subsection{General Analysis}\label{MassGap}

If the macroscopic black hole of the quantum theory, closely resembles the classical black hole configuration, the area and the mass are related through,
\begin{align}\label{I_R}
\frac{\mathcal{B} \I}{4\pi}=r_{\rm h}^{2}+a^{2}
=2M^{2}\left[1+\sqrt{1-\left\{\left(\frac{a}{M}\right)^{2}+\left(\frac{Q}{M}\right)^{2}\right\}}\right]-Q^{2}~.
\end{align}
It is possible to invert the above relation in order to write down the mass in terms of the discretized area as
\begin{align}
2M = \frac{\left(\mathcal{B} \I/4\pi\right)+Q^{2}}{\sqrt{\left(\mathcal{B} \I/4\pi\right)-a^{2}}}
= \sqrt{\left(\mathcal{B} \I/4\pi\right)-a^{2}}+\frac{\left(a^{2}+Q^{2}\right)}{\sqrt{\left(\mathcal{B} \I/4\pi\right)-a^{2}}}
\end{align}
We are interested in obtaining the smallest mass difference between two (mass-wise) nearest black holes. In principle, the nearest neighbour of a black hole of mass M (and hence integer $\I$) can only be larger than the mass identified by the integer value $\I+1$. In general, as both the mass and the rotation parameter of black hole are changed, the area of the black hole can either increase or decrease. For example, if the mass does not change, but only the rotation parameter decreases, the area will increase. However, for emission of a particle with certain energy, the mass parameter must be modified. Thus in general there will be a competition between the change in mass and the change in rotation parameter. For black holes which is near extremal it is difficult to argue, which one among these two will dominate. But for black holes having $a\ll M$, of course the change in mass will dominate and one can safely ignore the effects of rotation. A similar consideration applies to the charge parameter associated with the black hole as well. Since we are interested in black holes which are far from being extremal to start with, the area will decrease as it emits quanta of radiation.

\begin{figure*}
\begin{center}

\includegraphics[height=2in, width=3in]{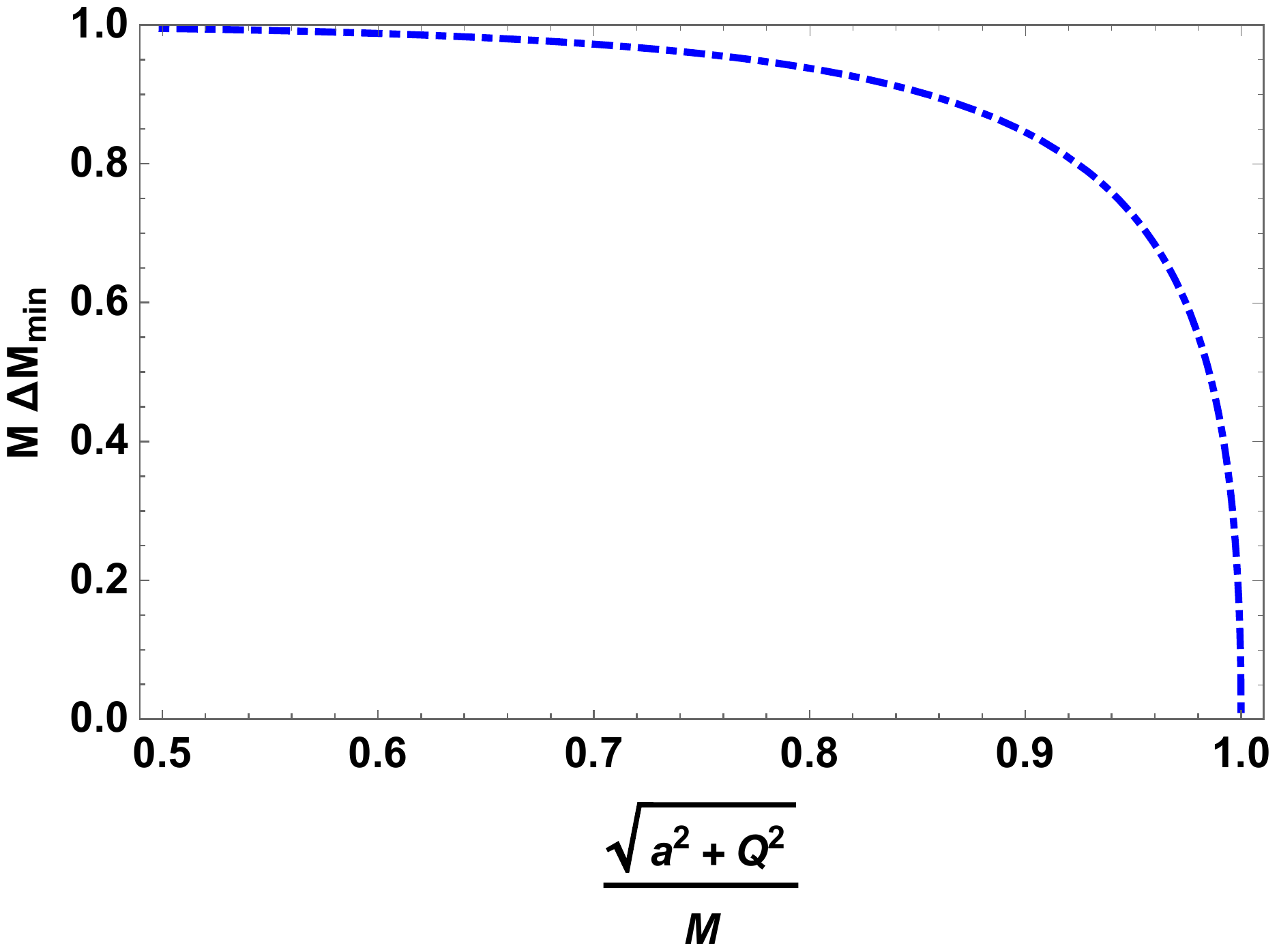}~~
\includegraphics[height=2in, width=3in]{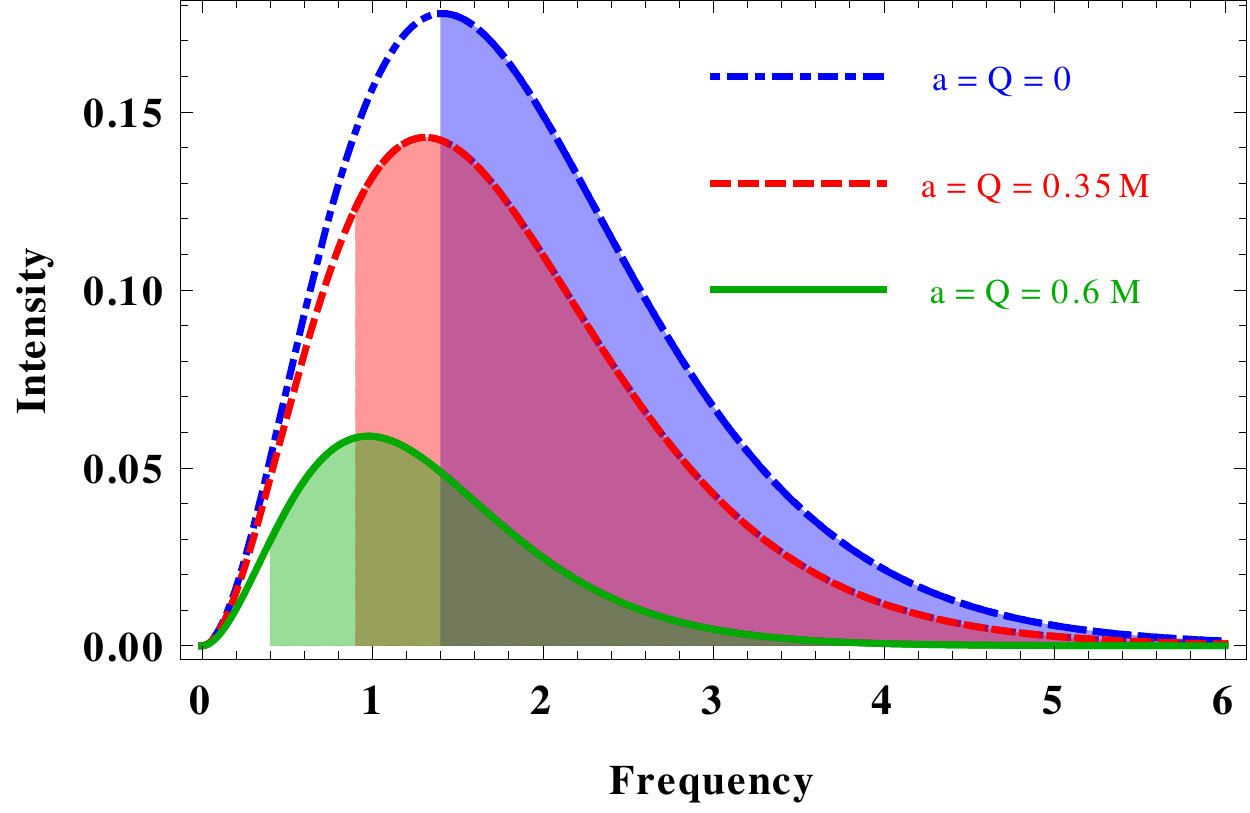}\\

\caption{The figure on the left depicts how the minimum mass gap, expressed for convenience as $M\Delta M_{\rm min}$, changes as $(\sqrt{a^{2}+Q^{2}}/M)$ approaches unity.  As the figure clearly demonstrates, with the black hole approaching extremal limit, the mass gap also tends to vanish. On the other hand, the right hand figure schematically illustrates how the quantum spectrum gradually becomes continuous as the black hole approaches the extremal limit. The blue, dot-dashed curve at the top illustrates the case of Schwarzschild black hole (i.e., with $a=Q=0$), from which it is clear that the spectrum will be largely discrete, as the minimum mass gap appears around the maxima of the distribution. While the red, dashed curve depicts the corresponding situation for $a=Q=0.35~M$ and the lowermost green curve illustrates the case with $a=Q=0.6~M$ respectively. As evident from the red and green curves, the region spanned by the dense spectrum gradually increases as the $(a/M)$ and $(Q/M)$ increases. Ultimately, when the black hole becomes near-extremal the quantum spectrum will become continuous.}\label{fig_01}
\end{center}
\end{figure*}

Following the above discussion we will be considering \emph{only} those transitions for which both the area and rotation parameter \emph{decrease}, since these are the ones which are relevant for a black hole away from extremality. Furthermore, we will be excluding transitions among black hole micro-states having identical area. Thus it follows that the mass change will be minimum when both $Q$ and $a$ are kept the same. Otherwise, the change in $Q$ or $a$ will add to the area change and hence the shift in black hole mass will be bigger. Thus the mass difference between two nearest 
micro-canonical configurations turns out to be
\begin{align}\label{MGap}
\Delta M _{\rm min}=\frac{\mathcal{B}}{16\pi} \frac{\left(\mathcal{B} \I/4\pi\right)-2a^{2}-Q^{2}}{\Big\{ 
\left(\mathcal{B} \I/4\pi\right)-a^{2}\Big\}^{3/2}}~.
\end{align}
In order to arrive at the above expression we have assumed the black hole to be macroscopic, which justifies expansion in inverse powers of $\mathcal{I}$. This is because, large value of $\mathcal{I}$ implies that most of the micro-states are already occupied, resulting into a macroscopic description for black holes. However, the above expression is not very illuminating. To cast the above relation to a more useful form we have to provide an expression for $(\mathcal{B} \I/4\pi)$ in terms of the hairs of the black hole. Such a relation is being supplied by \ref{I_R}, yielding,
\begin{align}
\frac{\mathcal{B} \I}{4\pi}-a^{2}=\left[M+\sqrt{M^{2}-\left(a^{2}+Q^{2}\right)}\right]^{2}~,\label{I_R2}\\
\frac{\mathcal{B} \I}{4\pi}-2a^{2}-Q^{2}=r_{\rm h}^{2}-\left(a^{2}+Q^{2}\right)
=2r_{\rm h}\sqrt{M^{2}-\left(a^{2}+Q^{2}\right)}, \label{I_R3}
\end{align}
which along with \ref{MGap} leads us to the relation,
\begin{align} \label{MGap2}
\Delta M _{\rm min}&=\frac{\mathcal{B}}{8\pi}\frac{\sqrt{M^{2}-\left(a^{2}+Q^{2}\right)}}{\left[M+\sqrt{M^{2}-\left(a^{2}+Q^{2}\right)}\right]^{2}}
=\frac{\mathcal{B}}{2}\frac{T_H(\eta)}{1+\sqrt{1-\eta ^{2}}}~.
\end{align}
Here $T_H(\eta)$ corresponds to the Hawking temperature associated with the Kerr-Newman black hole and $\eta ^{2}=(a^{2}+Q^{2})/M^{2}$, a dimensionless parameter approaching unity as the black hole approaches the extremal limit. The above expression clearly demonstrates that the minimum mass gap identically vanishes in the extremal limit. This is because, the term in the denominator takes a value $2$ in the $\eta \rightarrow 0$ limit, while it becomes $1$ in the extremal limit. On the other hand, the black hole temperature $T_H(\eta)$ is directly proportional to $1-\eta ^{2}$ and hence identically vanishes in the extremal limit. Therefore the minimum mass gap also vanishes, implying continuous black hole spectrum in the extremal limit as shown in \ref{fig_01}. One can further verify that this minimum mass gap exactly coincides with the one derived from thermodynamical consideration, a crucial test for consistency of the scenario depicted here.
We see that the (in principle) smallest mass gap gets related to the thermodynamic temperature of the black hole. If this mass gap gives rise to the emission of a quanta in the form of loss of energy, the frequency of the emitted quanta (ignoring the $\mathcal{B}/2$ factor), will always be less compared to the thermal frequency associated with black hole temperature. This is because the term $1+\sqrt{1-\eta ^{2}}$ is always greater than unity. Notably as, $a=Q=0$, then $\Delta M\sim 1/M$, as it should for a macroscopic Schwarzschild black hole \cite{Bekenstein:1995ju,Lochan:2015bha,Chakraborty:2016fye}. We would also like to point out that as the black hole approaches the classical extremal limit, i.e., $M\rightarrow (a^{2}+Q^{2})$, $\Delta M \sim 0$, remarkably the mass gap vanishes, leading to continuous quantum spectrum of emission, though at extremely small temperature. It turns out that continuous emission from rotating black hole in the extremal limit can also be derived from calculation of characteristic time scale of the emission of a quanta from black hole, see \cite{Hod:2015wva}.

The last property, presented above, is very unique for Kerr-Newman class of black holes and is absent in Schwarzschild black holes. The result clearly shows that near the extremal limit the quantum spectrum of a geometry discretized hole is almost continuous, unlike the Schwarzschild scenario. Thus the initial discreteness in the quantum spectrum of a Kerr-Newman black hole gradually makes its way to a continuous spectrum as the black hole approaches extremality (see also \cite{Hod:2015wva}).

Before concluding this discussion, let us briefly mention what happens to the mass gap in a more general scenario, in which both the mass and rotation parameter $a$ change. This is because we normally expect a black hole to emit its multi-pole moments which apart from carrying energy will carry away some angular momentum as well. In this situation both area and rotation parameter change once the black hole makes a jump to the nearest allowed configuration. That immediately leads to the following most general expression for the mass gap,
\begin{align}
\Delta M=\frac{\mathcal{B}}{8\pi}\frac{\sqrt{M^{2}-a^{2}}}{\left[M+\sqrt{M^{2}-a^{2}}\right]^{2}}
+\frac{Ma \Delta a}{\left[M+\sqrt{M^{2}-a^{2}}\right]^{2}}~.
\end{align}
 As evident from the above relation, by setting $\Delta a=0$ we get back the original relation for the minimum mass gap derived in \ref{MGap2} (with $Q=0$). This explicitly shows that the mass gap due to quantum nature of underlying geometry does not betray black hole thermodynamics, since it directly follows from the first law of black hole thermodynamics. Hence our initial assumption that black holes are macroscopic, behaving as a thermodynamic object holds good even when the geometry is discretized (see also \cite{Bekenstein:1997bt,Bekenstein:1995ju,Hod:1998vk}). Also note that for nonzero $\Delta a$ the mass gap will be higher from the minimum value. In what follows, we will concentrate on this characteristic minimum energy that a quantized (yet, macroscopic) black hole may emit and whether using this distinctive signature originating from the interface of quantum theory and gravity one can make any concrete prediction which can be falsified.
\subsection{Connecting micro and macro length scales}\label{Universal}

In order to understand the behaviour of the mass gap, let us now introduce two more length scales in the problem --- (a) the thermal de Broglie wavelength (for massless particles) $\lambda _{\rm T}\sim (1/T)$, signifying the scale set by the thermality of the black hole \cite{Barcelo:2010pj,Barcelo:2007yk,Visser:2001kq} and (b) the Compton wavelength $\lambda _{\rm c}\sim r_{\rm h}$ \cite{Bekenstein1974,Bekenstein:1997bt,Kotwal:2002ch,Hod:1998vk}, where $r_{\rm h}$ is the location of the event horizon, marking the {\it size} of the black hole. The smallest mass difference $\Delta M_{\rm min}$ corresponds to a minimum emission frequency and hence to a maximum emission wavelength $\lambda_{\rm max}$. Thus given \ref{MGap2}, the maximum wavelength takes the following form,
\begin{align}
\lambda _{\rm max}=\left(\frac{8\pi}{\mathcal{B}}\right)\frac{\left(M+\sqrt{M^{2}-\left(a^{2}+Q^{2}\right)}\right)^{2}}{\sqrt{M^{2}-\left(a^{2}+Q^{2}\right)}}~.
\end{align}
On the other hand, the thermal wavelength $\lambda _{\rm T}$ and the Compton wavelength $\lambda _{\rm c}$ for Kerr-Newman black hole take the following form,
\begin{align}
\lambda _{\rm T}&=\frac{4\pi M\left(M+\sqrt{M^{2}-\left(a^{2}+Q^{2}\right)}\right)}{\sqrt{M^{2}-\left(a^{2}+Q^{2}\right)}}~,
\label{thermal_W}
\\
\lambda _{\rm c}&=M+\sqrt{M^{2}-\left(a^{2}+Q^{2}\right)}~.
\label{comp_W}
\end{align}
It is possible to express the largest emission wavelength $\lambda _{\rm max}$ in terms of both $\lambda _{T}$ and $\lambda _{c}$ individually, i.e.,
\begin{align}
\lambda _{\rm max}&=\left(\frac{8\pi}{\mathcal{B}}\right)\left(\frac{\lambda _{\rm T}}{4\pi}\right)\left[1+\frac{1}{\left(\frac{\lambda _{\rm T}}{4\pi}\right)\frac{1}{M}-1}\right]~,
\end{align}
and
\begin{align}
\frac{1}{\lambda _{\rm max}}&=\frac{\mathcal{B}}{8\pi}\left(\frac{1}{\lambda _{\rm c}}-\frac{M}{\lambda _{\rm c}^{2}}\right).
\end{align}
Hence, we readily obtain that in the near-extremal limit, i.e., when $\left(a^{2}+Q^{2}\right)\rightarrow M^{2}$ and hence $\lambda _{\rm max}\sim \lambda _{\rm T}$, but \emph{not} as $\lambda _{\rm c}$. Thus for nearly extremal black holes the maximum wavelength (or, minimum frequency) scales with the thermal wavelength but \emph{not} the Compton wavelength.  Since the minimum emission frequency is intimately connected to the quasi-normal mode frequency of a black hole \cite{Dreyer:2002vy,Berti:2009kk,Chakraborty:2017qve}, the above observation suggests that in the near extremal limit, the minimum quasi-normal mode frequency of a black hole, has nothing to do with its size (which is $\sim M$) but with the thermal scale (which scales as $\sim \sqrt{M^{2}-\left(a^{2}+Q^{2}\right)}$)! Moreover, interestingly, it is possible to combine all the 
three length scales in a convenient manner to obtain,
\begin{align}\label{ScaleRel1}
\lambda _{\rm max}=\left(\frac{8\pi}{\mathcal{B}}\right)\frac{\lambda _{\rm c}\lambda _{\rm T}}{4\pi M}~. 
\end{align}
This expression relates the three distinct length scales of a black hole through its mass. Note that in the limit of vanishing charge and rotation, the thermal and the Compton wavelength both scale proportional to the black hole mass, such that the maximum wavelength for Schwarzschild black hole becomes $\lambda _{\rm max,sch}=(8\pi/\mathcal{B})4M$. Therefore, using the maximum wavelength for Schwarzschild, one readily arrives at the following dimensionless ratio,
\begin{align}\label{Scale_Rel2}
\frac{\lambda _{\rm max}}{\lambda _{\rm max,sch}}=\frac{\lambda _{\rm c}\lambda _{\rm T}}{16\pi M^{2}}~.
\end{align}
Since the maximum wavelength $\lambda _{\rm max}$ dictates whether the quantum spectrum of a black hole is dense or not, the above relation is a measure of the denseness of the emission spectral profile of a Kerr-Newman black hole viz-\'{a}-viz its Schwarzschild counterpart. One can actually write down a constant, independent of the nature of black hole hairs except its mass, out of the above length scales. This constant may be dubbed as universal among all the black holes having identical mass, i.e., it is independent of the charge and/or rotation parameter of the black hole. This can be achieved by noting that $\lambda _{\rm c,sch}=2M$ and $\lambda _{\rm T,sch}=8\pi M$, thus we obtain
\begin{align}
\frac{\lambda _{\rm max}}{\lambda _{\rm c}\lambda_{\rm T}}=\frac{\lambda _{\rm max,sch}}{\lambda _{\rm c,sch}\lambda_{\rm T,sch}}~.
\end{align}
The universality of the ratio $(\lambda _{\rm max}/\lambda _{\rm T}\lambda_{\rm c})$ stems from the fact that it {\it is independent of whether the black hole is Kerr-Newman or Schwarzschild} and speaks about the fundamental length scale of quantization of geometry (see \ref{ScaleRel1}).  Further, this tri-ratio of infrared physics cooks up an ultra-violet invariant of the theory of the black holes. Since this gives the parameter strength of the underlying quantum theory of black holes, it is an observationally realizable infrared test on the models of quantum gravity. When compared to using sub-leading corrections to the area-entropy relation \cite{Sen:2012dw,Chakraborty:2016dwb,Bhattacharya:2017bpl}, this tri-ratio is much more favourable to decide in favour of a theory observationally. Measurement of $\lambda_{c}$ for a black hole will be feasible in near future as the Einstein telescope starts operating \cite{Sathyaprakash:2012jk} as it can determine the photon radius of the black hole to sufficient accuracy. While measurement of $\lambda _{\rm T}$ and $\lambda _{\rm max}$ is not going to happen for a real black hole in foreseeable future, but it is indeed possible for analogue systems. There one can simply check whether such a relation exists and whether it can directly tell us something useful about the underlying quantum structure without having to probe ``quantum gravity''. 
\section{How long can a astrophysical black hole live?} \label{Lifetime}

We have obtained the expression for the smallest frequency to be emitted from a black hole as long as the mass of the black hole is large compared to the Planck mass. Therefore, in principle, we can follow the footprints of black hole emission obtained in earlier sections, till the time it turns into a Planck mass object, where the approximation of large mass, would no longer remain applicable. Therefore, through these calculations, we could in principle evaluate how many quanta a black hole could radiate before it settles into such a Planck mass system. Since we are interested in astrophysical black holes,  the presence of the electric charge can be safely neglected and hence we will only concentrate on Kerr black hole in what follows.
\subsection{Number of emitted quanta}

In order to obtain an upper bound on the number of emitted quanta, we will assume that throughout its evaporation the mass change brought about in the hole is only through the  minimum mass change at each step. Thus difference between the initial $\I_{i}$ (which signifies the starting hole mass) and the final $\I_{f}$ (signifying the Planck mass hole) should yield the number of quanta emitted. Let the initial mass be $M_{i}$ with the rotation parameter being $a_{i}$. Similarly, for the final state we have $a_{f}$ representing the final rotation parameter and $M_{f}$ being the final mass. Let $\eta ^{2} =a^{2}/M^{2}$, then for initial and final state we immediately obtain,
\begin{align}
\frac{\mathcal{B}}{4\pi}\I_{i}&=2M_{i}^{2}\left[1+\sqrt{1-\eta _{i}^{2}}\right]~,
\label{eq01}
\\
\frac{\mathcal{B}}{4\pi}\I_{f}&=2M_{f}^{2}\left[1+\sqrt{1-\eta _{f}^{2}}\right]~.
\label{eq02}
\end{align}
Introducing the final black hole mass a fraction of the initial mass  $M_{f}=\mu M_{i}$, we obtain the (largest) number of emitted quanta for the process $M_i \rightarrow M_f$, for the integer discretized black hole as,
\begin{align}\label{Num_Quant}
N_{\rm if}=\frac{8\pi}{\mathcal{B}}M_{i}^{2}\left[\left(1-\mu ^{2}\right)+\left(\sqrt{1-\eta _{i}^{2}}-\mu ^{2}\sqrt{1-\eta _{f}^{2}}\right)
\right]~.
\end{align}
In general if a macroscopic black hole emits thermal radiation as prescribed by semiclassical physics \cite{Hawking:1974sw,Chakraborty:2015nwa,Singh:2014paa}, then the quanta emission for a Kerr black hole will go on till it either becomes extremal or Planck sized. At the extremal limit, the temperature of the black hole becomes vanishingly small and hence the minimum energy will also go to zero (see \ref{fig_01}). In such a scenario, if the final state of the hole turns out to be extremal, then, $\eta _{f}=1$, hence the number of quanta emitted would be given by, 
\begin{align}
N_{\rm extremal}=\frac{8\pi}{\mathcal{B}}M_{i}^{2}\left[\left(1-\mu ^{2}\right)+\sqrt{1-\eta _{i}^{2}}\right]~.
\end{align}
With the end state being an extremal black hole, the maximum number of emitted quanta would correspond to the situation when besides being extremal the final black hole also turns out to be Planck sized, such that $\mu \rightarrow 0$, in which case the maximum number of quanta would be,
\begin{align}
N^{\rm max}_{\rm extremal}\simeq \frac{8\pi}{\mathcal{B}}M_{i}^{2}\left[1+\sqrt{1-\eta _{i}^{2}}\right]=\I_i~.
\end{align}
While the minimum number of emitted quanta for the extremal case would correspond to the case $\mu \sim 1$, i.e., the black hole becomes extremal very quickly much before reaching the Planck size and hence,
\begin{align}
N_{\rm extremal}^{\rm min}=\frac{8\pi}{\mathcal{B}}M_{i}^{2}\sqrt{1-\eta _{i}^{2}}~.
\end{align}
Further it will be useful to ensure that, the black hole, by emission of these quanta, is indeed going towards extremality. This can be done by imposing the condition that the rotation parameter and the black hole mass $M$ are changing in such a manner that $\eta _{f}>\eta _{i}$. For example, if $a_{f}^{2}>a_{i}^{2}$, while $M_{f}<M_{i}$ then the above criteria will be trivially satisfied. Also, note that for $\eta _{f}>\eta _{i}$, it follows that $(1-\eta _{f}^{2})<(1-\eta _{i}^{2})$ and hence, we arrive at, the inequality, $(\I_{f}/\I_{i})<\mu ^{2}$. Since the black hole is supposed to decay to lower and lower $\I$ values, the above condition ensures that it will certainly be driven to extremality.

However, as the black hole tends to reach an extremal configuration, i.e., the mass and the rotation parameter are comparable with one another, there are many factors that comes into play. First of all, in this case even for a finite decrease of mass and rotation parameter, the area can increase. Further, the determination of minimum mass gap becomes non-trivial and hence cannot be performed in a general context. This complicates the scenario significantly. In the semi-classical context it is certainly possible to perform a numerical analysis of the perturbation equations for various spin fields and hence determine the final state of the black hole. This has been worked out in \cite{Page:1976df,Page:1976ki} using numerical techniques, whose results we briefly recall. First of all for generic situations, with all possible spin fields into account, it is expected that the rotating black hole will end up as a Schwarzschild one with spin-1/2 particles carrying away maximum energy. On the other hand, if all the emissions are through scalar fields alone, it follows that the end product will be a rotating black hole with $a/M\sim 0.555$ \cite{Chambers:1997ai}. In our case as well, we are interested in change in black hole area such that mass changes by the minimum amount $\Delta M_{\rm min}$. This is certainly the case for scalar particles, as they do not take away any angular momentum. Further, the assumption that the black hole is always away from extremality is also borne out by the fact that the end product of a rotating black hole will have $a/M\sim 0.555<1$ \cite{Chambers:1997ai}. Hence, semi-classical physics would suggest to substitute $\eta _{\rm f}=0.555$ in \ref{Num_Quant}. Thus the analysis presented above naturally adopts to the semiclassical emission rates as well.

Furthermore, note that in our analysis we have considered quantum emission from black holes, rather than emission of quanta through semiclassical processes. Factors like grey body will make the emission process less efficient, leading to enhancement in lifetimes. However, we, in this work are more interested in finding out the quantum steps allowed for a macroscopic non extremal black hole. Hence in this case rather than the grey body factors, the distribution and probability of emission of these quanta is more important. These are the factors, which have been used in \cite{Bekenstein:1995ju} to determine the time scale of a black hole as we will discuss in the next section. Moreover, the analysis presented above provides an order of magnitude estimate and is by no means exact. There can be several additional factors including some form of the grey body factors, which will modify the numerical estimations. However, the key result, namely the number of quanta scaling as $M^{2}$ will not be affected.  

Thus given the initial mass and the final mass, along with the charge and rotation parameters it is possible to obtain the maximum number of quanta a black hole could emit. Depending upon the spin ${\cal S}$ of the emitted quanta, we can associate the dimensionality $(2{\cal S}+1)^N$ to the emitted radiation's Hilbert space quantum theoretically. Since the characteristic emission time for a quanta is related to the frequency it is emitted at, we can calculate, dynamically how long a black hole will last before it emits the quanta to turn into a Planck star or a extremal remnant. We can visualize the black hole emissions as jumps within macroscopically distinguishable configurations which must be microscopically orthogonal too. In that case we can also find out the fastest time in which a black hole can vanish \cite{MARGOLUS1998188}. This will also give the fastest route for a black hole to emit all its information. This exercise we leave for a future computation. 
\subsection{Lifetime of a black hole}

Given the number of maximum possible quanta of emission we can estimate  {\it the maximum possible lifetime} of a black hole.  To this end, we will assume that all the emitted quanta have the minimum energy $\omega _{\rm min}$, so that one can introduce a characteristic time scale $\tau$, identical for all the emitted quanta. A natural way to determine the time scale $\tau$ is through the following equation for the mass loss rate \cite{Bekenstein:1995ju}
\begin{align}\label{Ch_time}
\frac{dM}{dt}=-\frac{2\omega _{\rm min}}{\tau}~. 
\end{align}
At this stage one assumes some form for the mass loss rate and hence determine $\tau$. Subsequently this time scale is coupled to the number of emitted quanta, leading to an estimation of the lifetime of a black hole. Since we are considering only the minimum energy quanta, it immediately follows that the time period computed here will provide an upper bound to the lifetime of a black hole.

To proceed further it is natural to assume that the above mass loss is due to thermal emission. From which it is possible to determine the mass loss rate in terms of the area and the temperature of a astrophysical Kerr black hole as
\begin{align}\label{Eq_mass_loss}
\frac{dM}{dt}&=-\left(\frac{2\pi ^{2}}{120}\right)\left(\textrm{Area}\right)~T^{4}
\nonumber
\\
&=\left(-\frac{1}{1920 \pi}\right)\frac{\left[M^{2}-a^{2}\right]^{2}}{M^{3}\left(M+\sqrt{M^{2}-a^{2}}\right)^{3}}~,
\end{align}
where we have used expressions for the horizon area $A$ and the black hole temperature $T$ in term of the black hole parameters $(M,a)$. Note that the minus sign in front of \ref{Eq_mass_loss} ensures that the mass decreases with time. Finally equating \ref{Eq_mass_loss} with the right hand side of \ref{Ch_time}, we obtain the characteristic time scale $\tau$ of emission of a minimum energy quanta to be,
\begin{align}\label{timescale}
\tau =480\mathcal{B} M^{3} ~\frac{\left(M+\sqrt{M^{2}-a^{2}}\right)}{\Big\{M^{2}-a^{2}\Big\}^{3/2}}~.
\end{align}
Note that in the case of Schwarzschild black hole (i.e., with $a=0$) the time scale becomes proportional to $M$ \cite{Bekenstein:1995ju}, while in the extremal limit, obtained by taking $a^{2}\rightarrow M^{2}$, $\tau$ diverges. This suggests that in the near extremal region it takes a large time to emit the minimum energy quanta. This is expected as in the extremal limit the temperature of the black hole vanishes. The total lifetime of a black hole can be obtained by multiplying the average time scale $\tau$ with the total number of emitted quanta. In principle it should be obtained by integrating over the total time period of the black hole and since $\tau$ diverges, it takes an infinite time to reach the extremal limit. On the other hand, if we follow the analysis of \cite{Chambers:1997ai} with a large number of scalar fields being present whose quanta are emitted from the black hole, then for the final state $a/M\sim 0.555$. Thus one can simply substitute the same in \ref{timescale} to obtain $\tau \sim 1528 \mathcal{B}M$. This is three times larger compared to the corresponding Schwarzschild scenario. Thus the time scale of emission of a single quanta for a rotating black hole is larger compared to a static black hole. While, if the black hole is far from being extremal, as a crude estimate we can take the number of emitted quanta to be $\sim M^{2}$ and time scale of emission of a individual quanta as $\sim M$, then the total lifetime of a black hole will be $\sim M^{3}$, which is certainly large. The feasibility of this scheme for an isolated primordial black hole survivability is an interesting point to dwell upon, which we defer for future.
\section{Extremal limit of a black hole} \label{ExtremalLimit}

In this section we would like to show that a few results derived in the context of extremal limit of a macroscopic black hole also hold even when the black hole is assumed to have discrete structure. In the context of macroscopic physics it has been demonstrated that, if one throws a charged particle inside a Kerr (or, Kerr-Newman) black hole then one can gradually arrive at an extremal black hole. We would like to show that the same is true for our case as well, i.e., if we consider the black hole characterized by mass $M$ and angular momentum $J\equiv Ma$ to transform into another black hole with parameters $(\bar{M},\bar{J})$ by jumping down the discrete geometry spectrum, it actually evolves to an extremal configuration. We further assume that the rotating black hole emits an quanta with the minimum energy $\omega _{\rm min}$ with respect to the asymptotic observers, such that its mass change of the black hole corresponds to $\Delta M=\omega _{\rm min}$ and the change in angular momentum being $\Delta J=(
J/M)\Delta M$, such that $\delta a\equiv \delta (J/M)=0$. Then one can construct the following dimensionless quantity,
\begin{align}
\Theta (M,a)=\frac{T_{\rm new}-T_{\rm old}}{T_{\rm old}}~,
\end{align}
where, $T_{\rm old}$ corresponds to a rotating black hole with parameters $(M,J)$, while $T_{\rm new}$ relates to the black hole with $(M-\Delta M,J-\Delta J)$. Assuming $\Delta M \ll M$, we have the dimensionless ratio $\Theta$ taking the following form,
\begin{align}
\Theta (M,a)=\frac{M\Delta M}{M^{2}-a^{2}}\left[\sqrt{1-\eta ^{2}}-\eta ^{2}\right];\qquad \eta =\frac{a}{M}~.
\end{align}
Thus note that for $\eta \geq 0.79$ (which is a solution of the algebraic equation $\eta^{4}+\eta^{2}-1=0$), the dimensionless quantity $\Theta<0$ \cite{Hod:2017dck}. Hence as the black hole emits the minimum quanta the $\eta$ gradually increases as $a$ is fixed, while mass $M$ decreases, thus one will eventually end up with a higher negative value of the $\Theta$ parameter. This suggests that one arrives at the extremal limit as the black hole loses mass gradually by jumping into lower quantum states. An interesting fact with the above expression is that for $\eta<0.79$, the temperature associated with the new black hole is actually greater than the temperature of the old black hole, which is the signal of domination of Schwarzschild character again. 

Let us now briefly mention how to connect this up with the physical process of lowering a charged particle gradually into a Kerr black hole. In particular we will consider a situation where a particle of mass $\mu$ and charge $q$ is being dropped slowly into a Kerr black hole. We will assume that this will make the black hole to absorb $n$ quanta, such that it jumps from initial configuration $\I \rightarrow \I+n$ while the angular momentum $J$ is unaltered. Thus the original black hole with mass $M$ and rotation parameter $a=J/M$  becomes another black hole with mass $M+\mathcal{E}$ and rotation parameter $J/(M+\mathcal{E})$. Here $\mathcal{E}$ corresponds to the energy contributed by the charged object. Then from \ref{I_R} we obtain, 
\begin{align}
\frac{\mathcal{B} n}{4\pi}&=2(M+\mathcal{E})\left[(M+\mathcal{E})+\sqrt{(M+\mathcal{E})^{2}-\frac{J^{2}}{(M+\mathcal{E})^{2}}-q^{2}}\right]-q^{2}
-2M\left[M+\sqrt{M^{2}-\frac{J^{2}}{M^{2}}}\right]
\nonumber
\\
&=-\frac{Mq^{2}}{\sqrt{M^{2}-a^{2}}}-q^{2}+2\mathcal{E}\left[M+\sqrt{M^{2}-a^{2}}\right]
\left[1+\frac{M}{M+\sqrt{M^{2}-a^{2}}}+\frac{M^{2}+a^{2}}{\sqrt{M^{2}-a^{2}}\left[M+\sqrt{M^{2}-a^{2}}\right]} \right]~.
\end{align}
The above expression must remain finite in the extremal limit as well, since we would like to impose the condition that if the initial black hole is extremal then the final black hole should also remain extremal at most (i.e., does not become a naked singularity). Thus the requirement of divergent terms in the above expression in the extremal limit forces the energy associated with the charged particle to be
\begin{align}
\mathcal{E}=\frac{q^{2}}{4M}~.
\end{align}
Incidentally, this exactly coincides with the universal minimum energy that can be supplied by a charged body and is also the energy supplied in extremal situation by a charged particle \cite{Hod:2017dck}. This result was derived earlier using the classical black hole picture, while here we have merely computed the changed in the energy levels due to lowering of this charged particle in the black hole. Note that if the particle had no charge, i.e., if $q=0$ then the above expression will require $\mathcal{E}$ to identically vanish. Further using the above expression for $\mathcal{E}$, in the extremal limit we obtain,
\begin{align}\label{Quant_charge}
\frac{\mathcal{B} n}{4\pi}=\frac{\Delta A}{4\pi}=-q^{2}+4\mathcal{E}M=0~.
\end{align}
Thus remarkably, the area does not change when the energy carried by the charged body coincides with $q^{2}/4M$, the universal minimum energy that a charged object can supply. Hence as the black hole approaches to the extremal limit, the classical universal features associated with them are borne out by our quantum schemes as well.  
\section{Generalization to Higher Dimensional Black Holes} \label{HighDim}

So far, our discussion was concentrated on Kerr-Newman solution in four spacetime dimensions alone. In this setting we have observed that in the extremal limit the black hole emission spectra becomes almost continuous. We can similarly ask if such characteristic features of a rotating black hole are retained in higher dimensions. 

Let us start by considering Kerr black hole in higher dimensions, characterized with a single rotation parameter $a$ \cite{Emparan:2008eg,Myers:1986un,Kanti:2004nr,Frolov:2007nt,gravitation}. In $D$ spacetime dimensions the horizon of such a rotating black hole is located at $r=r_{h}$, where $r_{h}$ is a solution of the following algebraic equation,
\begin{align}
r^{2}+a^{2}-Mr^{5-D}=0~.
\end{align}
In general obtaining a solution for $r_{h}$ for arbitrary $D$ is difficult, so we will consider only the case $D=5$. For a five dimensional spacetime the above algebraic equation can be readily solved resulting into the following horizon location: $r_{h}=\sqrt{M-a^{2}}$. Note that in this case the extremal limit is defined as $M\rightarrow a^{2}$. The area of the event horizon for the rotating black hole in five dimensions can be easily computed, resulting into $A= 2\pi^{2} r_{h}(r_{h}^{2}+a^{2})$. Thus if the black hole is discretized in integer steps (see \ref{GeometricOp}) then the following relation can be obtained,
\begin{align}
\frac{\mathcal{B} \I}{2\pi ^{2}}=M\sqrt{M-a^{2}}~.
\end{align}
We can compute the minimum mass change for the higher dimensional solution following exactly an identical procedure as in \ref{MassGap}. In particular we keep the rotation parameter $a$ fixed and then consider the mass change as the integer $\I$ drops to $(\I-1)$. This results into the following expression for minimum mass change,
\begin{align}
\Delta M_{\rm min}=\frac{\mathcal{B}}{\pi ^{2}}\frac{\sqrt{M-a^{2}}}{3M-2a^{2}}~.
\end{align}
Thus again the mass change vanishes in the extremal limit, which corresponds to $M=a^{2}$. Thus our conclusion remains unchanged as we consider a five dimensional rotating solution. 

For completeness let us also consider a charged black hole in five dimensions. One can immediately compute the location of the event horizon by setting $g^{rr}=0$. This in turn implies that the horizon location can be obtained by solving the following algebraic equation, $r^{4}-2Mr^{2}+Q^{2}=0$. The corresponding solution for the event horizon becomes
\begin{align}
r_{h}^{2}=M+\sqrt{M^{2}-Q^{2}}~.
\end{align}
Here $M\rightarrow Q$ corresponds to the extremal limit. In this context the horizon area is being given by $A=2\pi ^{2}r_{h}^{3}$, such that for integer discretization scheme, the above area-mass relation reduces to, 
\begin{align}
\mathcal{B} ^{2/3}\I^{2/3}=M+\sqrt{M^{2}-Q^{2}}~,
\end{align}
One can in principle invert the above relation and obtain the mass in terms of the integer $\I$. Then again considering the transition of the black hole from $\I$ to $\I-1$, the minimum mass separation can be immediately obtained as,
\begin{align}
\Delta M=-\frac{\mathcal{B} Q^{2}}{3}\left[M+\sqrt{M^{2}-Q^{2}}\right]^{-5/2}+\frac{\mathcal{B}}{3}\left[M+\sqrt{M^{2}-Q^{2}}\right]^{-1/2}~.
\end{align}
In this case as well, note that in the extremal limit, i.e., as $M\rightarrow Q$, one essentially obtains a vanishing contribution to $\Delta M$. Thus in the case of a higher dimensional charged black hole, in the extremal limit, the quantum spectrum becomes dense again. Thus one can expect that in higher dimensions too, in the extremal limit, all the black holes in the Kerr-Newman class do produce continuous quantum spectrum, visually contrasting them with the Schwarzschild case. 
\section{Conclusion} \label{Conclusion}

In this work, we have studied the emission pattern of a classical black hole which is a macroscopic realization of an underlying quantum gravity theory, where some geometric feature of spacetime geometry gets discretized. We have obtained the spacing between distinguishable macroscopic configurations such a hole is allowed to have, which has revealed many interesting features when applied to the Kerr-Newman class of black holes. The smallest frequency of emission gets related to the temperature of the hole through the de-Broglie thermal wavelength associated with the emission profile. This is unlike a normal blackbody where the minimum frequency of emission is determined by the size of the blackbody. The computations presented here have many significant implications involving black hole information release, life time, scrambling time etc. Below we provide the key findings of this work:
\begin{itemize}

\item {\it Generality of the Approach:} The approach of area quantization presented here is completely general and certainly encompasses the more familiar integer area quantization scheme. Note that unlike previous works in this direction, the geometric operator which is being quantized is taken to be arbitrary, it could be area but need not be so. Then, the scheme of quantization is also arbitrary and determined through the function $f(j)$. Interestingly, all the results presented in this work goes through for any choices of $f(j)$ and even in this general setting the quantization of area can be established. Hence, as and when any theory of quantum gravity provides us a quantization scheme for geometric operators, the results presented here will remain directly applicable.

\item {\it Smallest frequency of emission:} The emission profile of the black hole admits a smallest mass (or frequency) gap. This comes just from the micro canonical counting of the entropy and is independent of any quantum modelling whatsoever. If the area changes by integer steps, the minimum frequency a black hole can emit, while jumping to nearest allowed configuration, scales inversely proportional to the hole's mass for the Schwarzschild case. Therefore, the collection of black hole radiation looks vastly discrete given the fact that the semi-classical radiation is expected to be thermal with the temperature inversely related to the mass. Similarly, for the Kerr-Newman black holes too, the minimum frequency gap between the nearest allowed configuration is set by the temperature the black hole is at. 

\item {\it Dense spectrum and extremal limit:} For Kerr-Newman black hole it is possible to consider the extremal limit. In this case the minimum mass gap becomes vanishing. Thus unlike the Schwarzschild scenario in this case the quantum spectrum can become dense, as the rotation parameter and/or charge of the black hole increases. 

\item {\it Connecting micro-physics to macro-physics:} For a macroscopic black hole, much of the information about its inherent quantum nature gets washed out. For example, its entropy, mass, maximum number of quanta it could emit would not change. However there are also quantities which know about the effect the quantum theory brought into the macroscopic parameters of the hole. We identified three such quantities: mass gap, time scale and a relation among the length scales. The last one among these is of significant interest. It relates three length scales associated with a black hole in a universal fashion. Therefore observation of these associated quantities presumably will reveal the parameter (e.g. Immirizi parameter \cite{Ghosh:2004rq}, string tension \cite{Callan:1988hs}, non commutativity parameter of space-time \cite{Nicolini:2005vd} etc.) associated with the black holes' discretization scheme.
 
\item {\it A road to Planck scale remnant:} Based upon our estimates of the minimum frequency of emission, we can calculate {\it maximum possible} number of quanta a black hole can emit before it turns extremal and/or becomes Planck sized. One can determine the number of quanta starting from the initial configuration, upto the time where the classical/semiclassical approximation breaks down. Therefore, one can also estimate the maximum possible information content, available with the radiation, once the evaporation has ended. With such emission profile one can also obtain a characteristic time scale of the black hole, which tells us about the time phase corresponding to a particular configuration a black hole adopts in course of emitting radiation. These studies may potentially provide insights regarding at least two crucial issues --- (a) a qualitative estimate of the amount of information a black hole is entitled to emit with its correspondence with information loss paradox and (b) an estimate for the lifetime of a  black hole, i.e., survivability of a black hole.
  
\end{itemize}

The results obtained here open up many frontiers of analysis, some of which are already discussed. We will like to pursue these aspects elsewhere. As the most useful insight obtained from this work, we would like to stress here is the infrared region of the black hole emission may be much more richer in terms of revealing the ultraviolet physics a microscopic theory is built up of. Further research in this direction is indeed warranted.
\section*{Acknowledgements}

Research of SC is funded by the INSPIRE Faculty Fellowship (Reg. No. DST/INSPIRE/04/2018/000893) from Department of Science and Technology, Government of India. Research of KL is supported by INSPIRE Faculty Fellowship grant by Department of Science and Technology, Government of India. 
\appendix
\labelformat{section}{Appendix #1} 
\labelformat{subsection}{Appendix #1}
\section{Appendix: Quantum gravitational correction to the area relation} \label{AppB}

From various quantum gravity models, the area entropy relation are expected to receive sub-dominant corrections of quantum gravitational origin \cite{Lochan:2012in,Lochan:2012sp,Sen:2012dw,Ghosh:2004rq}
of the form
\bea
S = \frac{A}{4 l_p^2}  + \frac{p}{q}\ln{\frac{\sigma A}{4 l_p^2}}~, \label{ModifiedRel}
\eea
where $p$ and $q$ are some integers and $l_{p}$ corresponds to the Planck length, where as $\sigma$ is a constant. Using the expression for area in terms of discretized geometrical operator, it follows that the above expression can be written as,
\bea
S=\frac{\mathcal{B}}{4 l_{p}^{2}}F^{\gamma}+\frac{p}{q}\ln{\frac{\sigma \mathcal{B}}{4 l_p^2}F^{\gamma}}~,
\eea
where the results from \cite{Lochan:2015bha} has been used. So far we have based our analysis on the leading order entropy-area relation. However in view of the above correction it is legitimate to ask the effect of the Logarithmic correction of area on the discrete quantum spectrum. We will show that such corrections terms will restrain the discretization scheme significantly and will also lead to more sparse emission spectrum. The number of microstates associated with the expression for entropy as presented in \ref{ModifiedRel} can be expressed as
\bea
\ln{g(n)}= \frac{\mathcal{B}}{4 l_p^2}F^{\gamma}+\frac{p}{q}\ln{\frac{\sigma \mathcal{B}}{4 l_p^2}F^{\gamma}}~.
\eea
Therefore, it immediately follows from the above expression that,
\bea
g(n)^q = F^{p\gamma}\exp{\tilde{\mathcal{B}}F^{\gamma}}~,
\eea
where we have absorbed the constants $\sigma$ and $\mathcal{B}/4l_{p}^{2}\sigma^{\gamma}$ in the redefinition of $F^{\gamma}$ and in defining a new parameter $\tilde{\mathcal{B}}$. Clearly for $g(n)$ to be an integer with $p,q \in \mathbb{Z}_+$, one requires $\tilde{\mathcal{B}} = \ln{\mathcal{I}_0}$ and $F^{\gamma}=\mathcal{I}$ to be the only possibility, where $\mathcal{I}$ and $\mathcal{I}_{0}$ are integers. Therefore the discussion with the integer step area variation as presented in \ref{MassGap} will dictate the emission profile. For the case $p<0$ one can obtain a superset of solution space from
\bea
g(n)^q = F^{-|p|\gamma}\exp{\tilde{\mathcal{B}} F^{\gamma}}~,
\eea
again leading to the same result as above, though with the use of a further condition: $\mathcal{I} = n \mathcal{I}_0$ for some integer $n$. Thus quantum gravity corrections in micro-canonical counting scheme will also lead to a largely discretize the black hole emission spectra.
\bibliography{Discrete_Rotating}

\providecommand{\href}[2]{#2}\begingroup\raggedright\begin{thebibliography}{10}

\bibitem{Bekenstein:1974ax}
J.~D. Bekenstein, ``{Generalized second law of thermodynamics in black hole
  physics},''
\href{http://dx.doi.org/10.1103/PhysRevD.9.3292}{{\em Phys. Rev.} {\bfseries
  D9} (1974) 3292--3300}.

\bibitem{Bardeen:1973gs}
J.~M. Bardeen, B.~Carter, and S.~W. Hawking, ``{The Four laws of black hole
  mechanics},''
\href{http://dx.doi.org/10.1007/BF01645742}{{\em Commun. Math. Phys.}
  {\bfseries 31} (1973) 161--170}.

\bibitem{Gibbons:1977mu}
G.~W. Gibbons and S.~W. Hawking, ``{Cosmological Event Horizons,
  Thermodynamics, and Particle Creation},''
\href{http://dx.doi.org/10.1103/PhysRevD.15.2738}{{\em Phys. Rev.} {\bfseries
  D15} (1977) 2738--2751}.

\bibitem{Hawking:1976de}
S.~W. Hawking, ``{Black Holes and Thermodynamics},''
\href{http://dx.doi.org/10.1103/PhysRevD.13.191}{{\em Phys. Rev.} {\bfseries
  D13} (1976) 191--197}.

\bibitem{Bekenstein:1973ur}
J.~D. Bekenstein, ``{Black holes and entropy},''
\href{http://dx.doi.org/10.1103/PhysRevD.7.2333}{{\em Phys. Rev.} {\bfseries
  D7} (1973) 2333--2346}.

\bibitem{Bekenstein:1972tm}
J.~D. Bekenstein, ``{Black holes and the second law},''
\href{http://dx.doi.org/10.1007/BF02757029}{{\em Lett. Nuovo Cim.} {\bfseries
  4} (1972) 737--740}.

\bibitem{Hawking:1974sw}
S.~W. Hawking, ``{Particle Creation by Black Holes},''
  \href{http://dx.doi.org/10.1007/BF02345020}{{\em Commun. Math. Phys.}
  {\bfseries 43} (1975) 199--220}.
[erratum,ibid,167(1975)].

\bibitem{Jacobson:1995ab}
T.~Jacobson, ``{Thermodynamics of space-time: The Einstein equation of
  state},'' \href{http://dx.doi.org/10.1103/PhysRevLett.75.1260}{{\em
  Phys.Rev.Lett.} {\bfseries 75} (1995) 1260--1263},
\href{http://arxiv.org/abs/gr-qc/9504004}{{\ttfamily arXiv:gr-qc/9504004
  [gr-qc]}}.

\bibitem{Padmanabhan:2009vy}
T.~Padmanabhan, ``{Thermodynamical Aspects of Gravity: New insights},''
  \href{http://dx.doi.org/10.1088/0034-4885/73/4/046901}{{\em Rept. Prog.
  Phys.} {\bfseries 73} (2010) 046901},
\href{http://arxiv.org/abs/0911.5004}{{\ttfamily arXiv:0911.5004 [gr-qc]}}.

\bibitem{Padmanabhan:2013nxa}
T.~Padmanabhan, ``{General Relativity from a Thermodynamic Perspective},'' {\em
  Gen.Rel.Grav.} {\bfseries 46} (2014) 1673,
\href{http://arxiv.org/abs/1312.3253}{{\ttfamily arXiv:1312.3253 [gr-qc]}}.

\bibitem{Maldacena:1996ky}
J.~M. Maldacena, {\em {Black holes in string theory}}.
\newblock PhD thesis, Princeton U., 1996.
\newblock \href{http://arxiv.org/abs/hep-th/9607235}{{\ttfamily
  arXiv:hep-th/9607235 [hep-th]}}.
\newblock
\url{http://wwwlib.umi.com/dissertations/fullcit?p9627605}.
\newblock

\bibitem{Rovelli:1996dv}
C.~Rovelli, ``{Black hole entropy from loop quantum gravity},''
  \href{http://dx.doi.org/10.1103/PhysRevLett.77.3288}{{\em Phys. Rev. Lett.}
  {\bfseries 77} (1996) 3288--3291},
\href{http://arxiv.org/abs/gr-qc/9603063}{{\ttfamily arXiv:gr-qc/9603063
  [gr-qc]}}.

\bibitem{Dowker:2005tz}
F.~Dowker, \href{http://dx.doi.org/10.1142/9789812700988_0016}{``{Causal sets
  and the deep structure of spacetime},''} in {\em 100 Years Of Relativity:
  space-time structure: Einstein and beyond}, A.~Ashtekar, ed., pp.~445--464.
\newblock 2005.
\newblock
\href{http://arxiv.org/abs/gr-qc/0508109}{{\ttfamily arXiv:gr-qc/0508109
  [gr-qc]}}.
\newblock

\bibitem{Ashtekar:1997yu}
A.~Ashtekar, J.~Baez, A.~Corichi, and K.~Krasnov, ``{Quantum geometry and black
  hole entropy},'' \href{http://dx.doi.org/10.1103/PhysRevLett.80.904}{{\em
  Phys. Rev. Lett.} {\bfseries 80} (1998) 904--907},
\href{http://arxiv.org/abs/gr-qc/9710007}{{\ttfamily arXiv:gr-qc/9710007
  [gr-qc]}}.

\bibitem{Hawking:1976ra}
S.~W. Hawking, ``{Breakdown of Predictability in Gravitational Collapse},''
\href{http://dx.doi.org/10.1103/PhysRevD.14.2460}{{\em Phys. Rev.} {\bfseries
  D14} (1976) 2460--2473}.

\bibitem{Visser:2014ypa}
M.~Visser, ``{Thermality of the Hawking flux},''
  \href{http://dx.doi.org/10.1007/JHEP07(2015)009}{{\em JHEP} {\bfseries 07}
  (2015) 009},
\href{http://arxiv.org/abs/1409.7754}{{\ttfamily arXiv:1409.7754 [gr-qc]}}.

\bibitem{Mathur:1997wb}
S.~D. Mathur, ``{Emission rates, the correspondence principle and the
  information paradox},''
  \href{http://dx.doi.org/10.1016/S0550-3213(98)00336-8}{{\em Nucl. Phys.}
  {\bfseries B529} (1998) 295--320},
\href{http://arxiv.org/abs/hep-th/9706151}{{\ttfamily arXiv:hep-th/9706151
  [hep-th]}}.

\bibitem{Mathur:2009hf}
S.~D. Mathur, ``{The Information paradox: A Pedagogical introduction},''
  \href{http://dx.doi.org/10.1088/0264-9381/26/22/224001}{{\em Class. Quant.
  Grav.} {\bfseries 26} (2009) 224001},
\href{http://arxiv.org/abs/0909.1038}{{\ttfamily arXiv:0909.1038 [hep-th]}}.

\bibitem{Hooft:2015jea}
G.~'t~Hooft, ``{Diagonalizing the Black Hole Information Retrieval Process},''
\href{http://arxiv.org/abs/1509.01695}{{\ttfamily arXiv:1509.01695 [gr-qc]}}.

\bibitem{Chakraborty:2017pmn}
S.~Chakraborty and K.~Lochan, ``{Black Holes: Eliminating Information or
  Illuminating New Physics?},''
  \href{http://dx.doi.org/10.3390/universe3030055}{{\em Universe} {\bfseries 3}
  no.~3, (2017) 55},
\href{http://arxiv.org/abs/1702.07487}{{\ttfamily arXiv:1702.07487 [gr-qc]}}.

\bibitem{Hawking:2016msc}
S.~W. Hawking, M.~J. Perry, and A.~Strominger, ``{Soft Hair on Black Holes},''
  \href{http://dx.doi.org/10.1103/PhysRevLett.116.231301}{{\em Phys. Rev.
  Lett.} {\bfseries 116} no.~23, (2016) 231301},
\href{http://arxiv.org/abs/1601.00921}{{\ttfamily arXiv:1601.00921 [hep-th]}}.

\bibitem{Hawking:2016sgy}
S.~W. Hawking, M.~J. Perry, and A.~Strominger, ``{Superrotation Charge and
  Supertranslation Hair on Black Holes},''
\href{http://arxiv.org/abs/1611.09175}{{\ttfamily arXiv:1611.09175 [hep-th]}}.

\bibitem{Modak:2014qja}
S.~K. Modak, L.~Ortíz, I.~Peña, and D.~Sudarsky, ``{Black hole evaporation:
  information loss but no paradox},''
  \href{http://dx.doi.org/10.1007/s10714-015-1960-y}{{\em Gen. Rel. Grav.}
  {\bfseries 47} no.~10, (2015) 120},
\href{http://arxiv.org/abs/1406.4898}{{\ttfamily arXiv:1406.4898 [gr-qc]}}.

\bibitem{Abbott:2016blz}
{\bfseries Virgo, LIGO Scientific} Collaboration, B.~P. Abbott {\em et~al.},
  ``{Observation of Gravitational Waves from a Binary Black Hole Merger},''
  \href{http://dx.doi.org/10.1103/PhysRevLett.116.061102}{{\em Phys. Rev.
  Lett.} {\bfseries 116} no.~6, (2016) 061102},
\href{http://arxiv.org/abs/1602.03837}{{\ttfamily arXiv:1602.03837 [gr-qc]}}.

\bibitem{Abbott:2016nmj}
{\bfseries Virgo, LIGO Scientific} Collaboration, B.~P. Abbott {\em et~al.},
  ``{GW151226: Observation of Gravitational Waves from a 22-Solar-Mass Binary
  Black Hole Coalescence},''
  \href{http://dx.doi.org/10.1103/PhysRevLett.116.241103}{{\em Phys. Rev.
  Lett.} {\bfseries 116} no.~24, (2016) 241103},
\href{http://arxiv.org/abs/1606.04855}{{\ttfamily arXiv:1606.04855 [gr-qc]}}.

\bibitem{Kraus:2015zda}
P.~Kraus and S.~D. Mathur, ``{Nature abhors a horizon},''
  \href{http://dx.doi.org/10.1142/S0218271815430038}{{\em Int. J. Mod. Phys.}
  {\bfseries D24} no.~12, (2015) 1543003},
\href{http://arxiv.org/abs/1505.05078}{{\ttfamily arXiv:1505.05078 [hep-th]}}.

\bibitem{Hawking:2014tga}
S.~W. Hawking, ``{Information Preservation and Weather Forecasting for Black
  Holes},''
\href{http://arxiv.org/abs/1401.5761}{{\ttfamily arXiv:1401.5761 [hep-th]}}.

\bibitem{Frolov:2014wja}
V.~P. Frolov, ``{Do Black Holes Exist?},'' in {\em {Proceedings, 18th
  International Seminar on High Energy Physics (Quarks 2014): Suzdal, Russia,
  June 2-8, 2014}}.
\newblock 2014.
\newblock \href{http://arxiv.org/abs/1411.6981}{{\ttfamily arXiv:1411.6981
  [hep-th]}}.
\newblock
\url{https://inspirehep.net/record/1329948/files/arXiv:1411.6981.pdf}.
\newblock

\bibitem{Vaz:2014era}
C.~Vaz, ``{Quantum gravitational dust collapse does not result in a black
  hole},'' \href{http://dx.doi.org/10.1016/j.nuclphysb.2014.12.021}{{\em Nucl.
  Phys.} {\bfseries B891} (2015) 558--569},
\href{http://arxiv.org/abs/1407.3823}{{\ttfamily arXiv:1407.3823 [gr-qc]}}.

\bibitem{Abedi}
J.~Abedi, H.~Dykaar, and N.~Afshordi, ``Echoes from the abyss: Tentative
  evidence for planck-scale structure at black hole horizons,''
  \href{http://dx.doi.org/10.1103/PhysRevD.96.082004}{{\em Phys. Rev. D}
  {\bfseries 96} (Oct, 2017) 082004}.
  \url{https://link.aps.org/doi/10.1103/PhysRevD.96.082004}.

\bibitem{Foit:2016uxn}
V.~F. Foit and M.~Kleban, ``{Testing Quantum Black Holes with Gravitational
  Waves},''
\href{http://arxiv.org/abs/1611.07009}{{\ttfamily arXiv:1611.07009 [hep-th]}}.

\bibitem{Cardoso:2017cqb}
V.~Cardoso and P.~Pani, ``{Tests for the existence of horizons through
  gravitational wave echoes},''
  \href{http://dx.doi.org/10.1038/s41550-017-0225-y}{{\em Nat. Astron.}
  {\bfseries 1} (2017) 586--591},
\href{http://arxiv.org/abs/1709.01525}{{\ttfamily arXiv:1709.01525 [gr-qc]}}.

\bibitem{Maselli:2017kic}
A.~Maselli, P.~Pani, R.~Cotesta, L.~Gualtieri, V.~Ferrari, and L.~Stella,
  ``{Geodesic models of quasi-periodic-oscillations as probes of quadratic
  gravity},'' \href{http://dx.doi.org/10.3847/1538-4357/aa72e2}{{\em Astrophys.
  J.} {\bfseries 843} no.~1, (2017) 25},
\href{http://arxiv.org/abs/1703.01472}{{\ttfamily arXiv:1703.01472
  [astro-ph.HE]}}.

\bibitem{Cardoso:2017cfl}
V.~Cardoso, E.~Franzin, A.~Maselli, P.~Pani, and G.~Raposo, ``{Testing
  strong-field gravity with tidal Love numbers},''
  \href{http://dx.doi.org/10.1103/PhysRevD.95.089901,
  10.1103/PhysRevD.95.084014}{{\em Phys. Rev.} {\bfseries D95} no.~8, (2017)
  084014}, \href{http://arxiv.org/abs/1701.01116}{{\ttfamily arXiv:1701.01116
  [gr-qc]}}.
[Addendum: Phys. Rev.D95,no.8,089901(2017)].

\bibitem{Alonso-Serrano:2015trn}
A.~Alonso-Serrano and M.~Visser, ``{On burning a lump of coal},''
  \href{http://dx.doi.org/10.1016/j.physletb.2016.04.023}{{\em Phys. Lett.}
  {\bfseries B757} (2016) 383--386},
\href{http://arxiv.org/abs/1511.01162}{{\ttfamily arXiv:1511.01162 [gr-qc]}}.

\bibitem{Lochan:2015oba}
K.~Lochan and T.~Padmanabhan, ``{Extracting information about the initial state
  from the black hole radiation},''
  \href{http://dx.doi.org/10.1103/PhysRevLett.116.051301}{{\em Phys. Rev.
  Lett.} {\bfseries 116} no.~5, (2016) 051301},
\href{http://arxiv.org/abs/1507.06402}{{\ttfamily arXiv:1507.06402 [gr-qc]}}.

\bibitem{Lochan:2016nbs}
K.~Lochan, S.~Chakraborty, and T.~Padmanabhan, ``{Information retrieval from
  black holes},'' \href{http://dx.doi.org/10.1103/PhysRevD.94.044056}{{\em
  Phys. Rev.} {\bfseries D94} no.~4, (2016) 044056},
\href{http://arxiv.org/abs/1604.04987}{{\ttfamily arXiv:1604.04987 [gr-qc]}}.

\bibitem{Chakraborty:2016fye}
S.~Chakraborty and K.~Lochan, ``{Quantum leaps of black holes: Magnifying
  glasses of quantum gravity},''
  \href{http://dx.doi.org/10.1142/S0218271816440247}{{\em Int. J. Mod. Phys.}
  {\bfseries D25} (2016) 1644024},
\href{http://arxiv.org/abs/1606.04348}{{\ttfamily arXiv:1606.04348 [gr-qc]}}.

\bibitem{Saini:2015dea}
A.~Saini and D.~Stojkovic, ``{Radiation from a collapsing object is manifestly
  unitary},'' \href{http://dx.doi.org/10.1103/PhysRevLett.114.111301}{{\em
  Phys. Rev. Lett.} {\bfseries 114} no.~11, (2015) 111301},
\href{http://arxiv.org/abs/1503.01487}{{\ttfamily arXiv:1503.01487 [gr-qc]}}.

\bibitem{Chen:2014jwq}
P.~Chen, Y.~C. Ong, and D.-h. Yeom, ``{Black Hole Remnants and the Information
  Loss Paradox},'' \href{http://dx.doi.org/10.1016/j.physrep.2015.10.007}{{\em
  Phys. Rept.} {\bfseries 603} (2015) 1--45},
\href{http://arxiv.org/abs/1412.8366}{{\ttfamily arXiv:1412.8366 [gr-qc]}}.

\bibitem{Bekenstein:1995ju}
J.~D. Bekenstein and V.~F. Mukhanov, ``{Spectroscopy of the quantum black
  hole},'' \href{http://dx.doi.org/10.1016/0370-2693(95)01148-J}{{\em Phys.
  Lett.} {\bfseries B360} (1995) 7--12},
\href{http://arxiv.org/abs/gr-qc/9505012}{{\ttfamily arXiv:gr-qc/9505012
  [gr-qc]}}.

\bibitem{Bekenstein1974}
J.~D. Bekenstein, ``The quantum mass spectrum of the kerr black hole,''
  \href{http://dx.doi.org/10.1007/BF02762768}{{\em Lettere al Nuovo Cimento
  (1971-1985)} {\bfseries 11} no.~9, (Nov, 1974) 467--470}.
  \url{https://doi.org/10.1007/BF02762768}.

\bibitem{Bekenstein:1997bt}
J.~D. Bekenstein, ``{Quantum black holes as atoms},'' in {\em {Recent
  developments in theoretical and experimental general relativity, gravitation,
  and relativistic field theories. Proceedings, 8th Marcel Grossmann meeting,
  MG8, Jerusalem, Israel, June 22-27, 1997. Pts. A, B}}, pp.~92--111.
\newblock 1997.
\newblock
\href{http://arxiv.org/abs/gr-qc/9710076}{{\ttfamily arXiv:gr-qc/9710076
  [gr-qc]}}.
\newblock

\bibitem{Lochan:2015bha}
K.~Lochan and S.~Chakraborty, ``{Discrete quantum spectrum of black holes},''
  \href{http://dx.doi.org/10.1016/j.physletb.2016.01.060}{{\em Phys. Lett.}
  {\bfseries B755} (2016) 37--42},
\href{http://arxiv.org/abs/1509.09010}{{\ttfamily arXiv:1509.09010 [gr-qc]}}.

\bibitem{Gray:2015pma}
F.~Gray, S.~Schuster, A.~Van–Brunt, and M.~Visser, ``{The Hawking cascade
  from a black hole is extremely sparse},''
  \href{http://dx.doi.org/10.1088/0264-9381/33/11/115003}{{\em Class. Quant.
  Grav.} {\bfseries 33} no.~11, (2016) 115003},
\href{http://arxiv.org/abs/1506.03975}{{\ttfamily arXiv:1506.03975 [gr-qc]}}.

\bibitem{Hod:2015wva}
S.~Hod, ``{The Hawking evaporation process of rapidly-rotating black holes: An
  almost continuous cascade of gravitons},''
  \href{http://dx.doi.org/10.1140/epjc/s10052-015-3554-y}{{\em Eur. Phys. J.}
  {\bfseries C75} no.~7, (2015) 329},
\href{http://arxiv.org/abs/1506.05457}{{\ttfamily arXiv:1506.05457 [gr-qc]}}.

\bibitem{Hod:2016xbu}
S.~Hod, ``{The entropy emission properties of near-extremal
  Reissner-Nordstr\"om black holes},''
  \href{http://dx.doi.org/10.1103/PhysRevD.93.104027}{{\em Phys. Rev.}
  {\bfseries D93} no.~10, (2016) 104027},
\href{http://arxiv.org/abs/1606.04944}{{\ttfamily arXiv:1606.04944 [gr-qc]}}.

\bibitem{Hod:2016rmg}
S.~Hod, ``{The Hawking cascades of gravitons from higher-dimensional
  Schwarzschild black holes},''
  \href{http://dx.doi.org/10.1016/j.physletb.2016.03.002}{{\em Phys. Lett.}
  {\bfseries B756} (2016) 133--136},
\href{http://arxiv.org/abs/1605.08440}{{\ttfamily arXiv:1605.08440 [gr-qc]}}.

\bibitem{Hod:1998vk}
S.~Hod, ``{Bohr's correspondence principle and the area spectrum of quantum
  black holes},'' \href{http://dx.doi.org/10.1103/PhysRevLett.81.4293}{{\em
  Phys. Rev. Lett.} {\bfseries 81} (1998) 4293},
\href{http://arxiv.org/abs/gr-qc/9812002}{{\ttfamily arXiv:gr-qc/9812002
  [gr-qc]}}.

\bibitem{Hod:2000it}
S.~Hod, ``{Gravitation, the quantum, and Bohr's correspondence principle},''
  \href{http://dx.doi.org/10.1023/A:1026753914838}{{\em Gen. Rel. Grav.}
  {\bfseries 31} (1999) 1639},
\href{http://arxiv.org/abs/gr-qc/0002002}{{\ttfamily arXiv:gr-qc/0002002
  [gr-qc]}}.

\bibitem{Barreira:1996dt}
M.~Barreira, M.~Carfora, and C.~Rovelli, ``{Physics with nonperturbative
  quantum gravity: Radiation from a quantum black hole},''
  \href{http://dx.doi.org/10.1007/BF02109521}{{\em Gen. Rel. Grav.} {\bfseries
  28} (1996) 1293--1299},
\href{http://arxiv.org/abs/gr-qc/9603064}{{\ttfamily arXiv:gr-qc/9603064
  [gr-qc]}}.

\bibitem{Lochan:2012in}
K.~Lochan and C.~Vaz, ``{Canonical Partition function of Loop Black Holes},''
  \href{http://dx.doi.org/10.1103/PhysRevD.85.104041}{{\em Phys. Rev.}
  {\bfseries D85} (2012) 104041},
\href{http://arxiv.org/abs/1202.2301}{{\ttfamily arXiv:1202.2301 [gr-qc]}}.

\bibitem{Lochan:2012sp}
K.~Lochan and C.~Vaz, ``{Statistical analysis of entropy correction from
  topological defects in Loop Black Holes},''
  \href{http://dx.doi.org/10.1103/PhysRevD.86.044035}{{\em Phys. Rev.}
  {\bfseries D86} (2012) 044035},
\href{http://arxiv.org/abs/1205.3974}{{\ttfamily arXiv:1205.3974 [gr-qc]}}.

\bibitem{Sen:2012dw}
A.~Sen, ``{Logarithmic Corrections to Schwarzschild and Other Non-extremal
  Black Hole Entropy in Different Dimensions},''
  \href{http://dx.doi.org/10.1007/JHEP04(2013)156}{{\em JHEP} {\bfseries 04}
  (2013) 156},
\href{http://arxiv.org/abs/1205.0971}{{\ttfamily arXiv:1205.0971 [hep-th]}}.

\bibitem{Ghosh:2004rq}
A.~Ghosh and P.~Mitra, ``{A Bound on the log correction to the black hole area
  law},'' \href{http://dx.doi.org/10.1103/PhysRevD.71.027502}{{\em Phys. Rev.}
  {\bfseries D71} (2005) 027502},
\href{http://arxiv.org/abs/gr-qc/0401070}{{\ttfamily arXiv:gr-qc/0401070
  [gr-qc]}}.

\bibitem{Rovelli:1994ge}
C.~Rovelli and L.~Smolin, ``{Discreteness of area and volume in quantum
  gravity},'' \href{http://dx.doi.org/10.1016/0550-3213(95)00150-Q,
  10.1016/0550-3213(95)00550-5}{{\em Nucl. Phys.} {\bfseries B442} (1995)
  593--622}, \href{http://arxiv.org/abs/gr-qc/9411005}{{\ttfamily
  arXiv:gr-qc/9411005 [gr-qc]}}.
[Erratum: Nucl. Phys.B456,753(1995)].

\bibitem{Vaz:2007td}
C.~Vaz, S.~Gutti, C.~Kiefer, T.~P. Singh, and L.~C.~R. Wijewardhana, ``{Mass
  spectrum and statistical entropy of the BTZ black hole from canonical quantum
  gravity},'' \href{http://dx.doi.org/10.1103/PhysRevD.77.064021}{{\em Phys.
  Rev.} {\bfseries D77} (2008) 064021},
\href{http://arxiv.org/abs/0712.1998}{{\ttfamily arXiv:0712.1998 [gr-qc]}}.

\bibitem{Vaz:2009jj}
C.~Vaz and L.~C.~R. Wijewardhana, ``{Spectrum and Entropy of AdS Black
  Holes},'' \href{http://dx.doi.org/10.1103/PhysRevD.79.084014}{{\em Phys.
  Rev.} {\bfseries D79} (2009) 084014},
\href{http://arxiv.org/abs/0902.1192}{{\ttfamily arXiv:0902.1192 [gr-qc]}}.

\bibitem{Sen2014}
A.~Sen, ``Microscopic and macroscopic entropy of extremal black holes in string
  theory,'' \href{http://dx.doi.org/10.1007/s10714-014-1711-5}{{\em General
  Relativity and Gravitation} {\bfseries 46} no.~5, (Apr, 2014) 1711}.
  \url{https://doi.org/10.1007/s10714-014-1711-5}.

\bibitem{Kaul:2012pf}
R.~K. Kaul, ``{Entropy of quantum black holes},'' {\em SIGMA} {\bfseries 8}
  (2012) 005,
\href{http://arxiv.org/abs/1201.6102}{{\ttfamily arXiv:1201.6102 [gr-qc]}}.

\bibitem{Visser:1992ck}
M.~Visser, ``{Hawking radiation: A Particle physics perspective},''
  \href{http://dx.doi.org/10.1142/S0217732393001409}{{\em Mod. Phys. Lett.}
  {\bfseries A8} (1993) 1661--1670},
\href{http://arxiv.org/abs/hep-th/9204062}{{\ttfamily arXiv:hep-th/9204062
  [hep-th]}}.

\bibitem{Rovelli:2017mzl}
C.~Rovelli, ``{Black holes have more states than those giving the
  Bekenstein-Hawking entropy: a simple argument},''
\href{http://arxiv.org/abs/1710.00218}{{\ttfamily arXiv:1710.00218 [gr-qc]}}.

\bibitem{Chakraborty:2017s}
S.~Chakraborty, \href{http://dx.doi.org/10.1007/978-3-319-63733-4}{{\em
  {Classical and Quantum Aspects of Gravity in Relation to The Emergent
  Paradigm}}}.
\newblock Springer, 2017.

\bibitem{Barcelo:2010pj}
C.~Barcelo, S.~Liberati, S.~Sonego, and M.~Visser, ``{Minimal conditions for
  the existence of a Hawking-like flux},''
  \href{http://dx.doi.org/10.1103/PhysRevD.83.041501}{{\em Phys. Rev.}
  {\bfseries D83} (2011) 041501},
\href{http://arxiv.org/abs/1011.5593}{{\ttfamily arXiv:1011.5593 [gr-qc]}}.

\bibitem{Barcelo:2007yk}
C.~Barcelo, S.~Liberati, S.~Sonego, and M.~Visser, ``{Fate of gravitational
  collapse in semiclassical gravity},''
  \href{http://dx.doi.org/10.1103/PhysRevD.77.044032}{{\em Phys. Rev.}
  {\bfseries D77} (2008) 044032},
\href{http://arxiv.org/abs/0712.1130}{{\ttfamily arXiv:0712.1130 [gr-qc]}}.

\bibitem{Visser:2001kq}
M.~Visser, ``{Essential and inessential features of Hawking radiation},''
  \href{http://dx.doi.org/10.1142/S0218271803003190}{{\em Int. J. Mod. Phys.}
  {\bfseries D12} (2003) 649--661},
\href{http://arxiv.org/abs/hep-th/0106111}{{\ttfamily arXiv:hep-th/0106111
  [hep-th]}}.

\bibitem{Kotwal:2002ch}
A.~V. Kotwal and S.~Hofmann, ``{Discrete energy spectrum of Hawking radiation
  from Schwarzschild surfaces},''
\href{http://arxiv.org/abs/hep-ph/0204117}{{\ttfamily arXiv:hep-ph/0204117
  [hep-ph]}}.

\bibitem{Dreyer:2002vy}
O.~Dreyer, ``{Quasinormal modes, the area spectrum, and black hole entropy},''
  \href{http://dx.doi.org/10.1103/PhysRevLett.90.081301}{{\em Phys. Rev. Lett.}
  {\bfseries 90} (2003) 081301},
\href{http://arxiv.org/abs/gr-qc/0211076}{{\ttfamily arXiv:gr-qc/0211076
  [gr-qc]}}.

\bibitem{Berti:2009kk}
E.~Berti, V.~Cardoso, and A.~O. Starinets, ``{Quasinormal modes of black holes
  and black branes},''
  \href{http://dx.doi.org/10.1088/0264-9381/26/16/163001}{{\em Class. Quant.
  Grav.} {\bfseries 26} (2009) 163001},
\href{http://arxiv.org/abs/0905.2975}{{\ttfamily arXiv:0905.2975 [gr-qc]}}.

\bibitem{Chakraborty:2017qve}
S.~Chakraborty, K.~Chakravarti, S.~Bose, and S.~SenGupta, ``{Signatures of
  extra dimensions in gravitational waves from black hole quasi-normal
  modes},''
\href{http://arxiv.org/abs/1710.05188}{{\ttfamily arXiv:1710.05188 [gr-qc]}}.

\bibitem{Chakraborty:2016dwb}
S.~Chakraborty, S.~Bhattacharya, and T.~Padmanabhan, ``{Entropy of a generic
  null surface from its associated Virasoro algebra},''
  \href{http://dx.doi.org/10.1016/j.physletb.2016.10.059}{{\em Phys. Lett.}
  {\bfseries B763} (2016) 347--351},
\href{http://arxiv.org/abs/1605.06988}{{\ttfamily arXiv:1605.06988 [gr-qc]}}.

\bibitem{Bhattacharya:2017bpl}
S.~Bhattacharya, S.~Chakraborty, and T.~Padmanabhan, ``{Entropy of a box of gas
  in an external gravitational field - revisited},''
  \href{http://dx.doi.org/10.1103/PhysRevD.96.084030}{{\em Phys. Rev.}
  {\bfseries D96} no.~8, (2017) 084030},
\href{http://arxiv.org/abs/1702.08723}{{\ttfamily arXiv:1702.08723 [gr-qc]}}.

\bibitem{Sathyaprakash:2012jk}
B.~Sathyaprakash {\em et~al.}, ``{Scientific Objectives of Einstein
  Telescope},'' \href{http://dx.doi.org/10.1088/0264-9381/29/12/124013,
  10.1088/0264-9381/30/7/079501}{{\em Class. Quant. Grav.} {\bfseries 29}
  (2012) 124013}, \href{http://arxiv.org/abs/1206.0331}{{\ttfamily
  arXiv:1206.0331 [gr-qc]}}.
[Erratum: Class. Quant. Grav.30,079501(2013)].

\bibitem{Chakraborty:2015nwa}
S.~Chakraborty, S.~Singh, and T.~Padmanabhan, ``{A quantum peek inside the
  black hole event horizon},''
  \href{http://dx.doi.org/10.1007/JHEP06(2015)192}{{\em JHEP} {\bfseries 06}
  (2015) 192},
\href{http://arxiv.org/abs/1503.01774}{{\ttfamily arXiv:1503.01774 [gr-qc]}}.

\bibitem{Singh:2014paa}
S.~Singh and S.~Chakraborty, ``{Black hole kinematics: The “in”-vacuum
  energy density and flux for different observers},''
  \href{http://dx.doi.org/10.1103/PhysRevD.90.024011}{{\em Phys. Rev.}
  {\bfseries D90} no.~2, (2014) 024011},
\href{http://arxiv.org/abs/1404.0684}{{\ttfamily arXiv:1404.0684 [gr-qc]}}.

\bibitem{Page:1976df}
D.~N. Page, ``{Particle Emission Rates from a Black Hole: Massless Particles
  from an Uncharged, Nonrotating Hole},''
\href{http://dx.doi.org/10.1103/PhysRevD.13.198}{{\em Phys. Rev.} {\bfseries
  D13} (1976) 198--206}.

\bibitem{Page:1976ki}
D.~N. Page, ``{Particle Emission Rates from a Black Hole. 2. Massless Particles
  from a Rotating Hole},''
\href{http://dx.doi.org/10.1103/PhysRevD.14.3260}{{\em Phys. Rev.} {\bfseries
  D14} (1976) 3260--3273}.

\bibitem{Chambers:1997ai}
C.~M. Chambers, W.~A. Hiscock, and B.~Taylor, ``{Spinning down a black hole
  with scalar fields},''
  \href{http://dx.doi.org/10.1103/PhysRevLett.78.3249}{{\em Phys. Rev. Lett.}
  {\bfseries 78} (1997) 3249--3251},
\href{http://arxiv.org/abs/gr-qc/9703018}{{\ttfamily arXiv:gr-qc/9703018
  [gr-qc]}}.

\bibitem{MARGOLUS1998188}
N.~Margolus and L.~B. Levitin, ``The maximum speed of dynamical evolution,''
  \href{http://dx.doi.org/https://doi.org/10.1016/S0167-2789(98)00054-2}{{\em
  Physica D: Nonlinear Phenomena} {\bfseries 120} no.~1, (1998) 188 -- 195}.
  \url{http://www.sciencedirect.com/science/article/pii/S0167278998000542}.
  Proceedings of the Fourth Workshop on Physics and Consumption.

\bibitem{Hod:2017dck}
S.~Hod, ``{A lower bound on the Bekenstein-Hawking temperature of black
  holes},'' \href{http://dx.doi.org/10.1016/j.physletb.2016.06.021}{{\em Phys.
  Lett.} {\bfseries B759} (2016) 541},
\href{http://arxiv.org/abs/1701.00492}{{\ttfamily arXiv:1701.00492 [gr-qc]}}.

\bibitem{Emparan:2008eg}
R.~Emparan and H.~S. Reall, ``{Black Holes in Higher Dimensions},''
  \href{http://dx.doi.org/10.12942/lrr-2008-6}{{\em Living Rev. Rel.}
  {\bfseries 11} (2008) 6},
\href{http://arxiv.org/abs/0801.3471}{{\ttfamily arXiv:0801.3471 [hep-th]}}.

\bibitem{Myers:1986un}
R.~C. Myers and M.~J. Perry, ``{Black Holes in Higher Dimensional
  Space-Times},''
\href{http://dx.doi.org/10.1016/0003-4916(86)90186-7}{{\em Annals Phys.}
  {\bfseries 172} (1986) 304}.

\bibitem{Kanti:2004nr}
P.~Kanti, ``{Black holes in theories with large extra dimensions: A Review},''
  \href{http://dx.doi.org/10.1142/S0217751X04018324}{{\em Int. J. Mod. Phys.}
  {\bfseries A19} (2004) 4899--4951},
\href{http://arxiv.org/abs/hep-ph/0402168}{{\ttfamily arXiv:hep-ph/0402168
  [hep-ph]}}.

\bibitem{Frolov:2007nt}
V.~P. Frolov and D.~Kubiznak, ``{Hidden Symmetries of Higher Dimensional
  Rotating Black Holes},''
  \href{http://dx.doi.org/10.1103/PhysRevLett.98.011101}{{\em Phys. Rev. Lett.}
  {\bfseries 98} (2007) 011101},
\href{http://arxiv.org/abs/gr-qc/0605058}{{\ttfamily arXiv:gr-qc/0605058
  [gr-qc]}}.

\bibitem{gravitation}
T.Padmanabhan, {\em {Gravitation: Foundations and Frontiers}}.
\newblock Cambridge University Press, Cambridge, UK, 2010.

\bibitem{Callan:1988hs}
C.~G. Callan, Jr., R.~C. Myers, and M.~J. Perry, ``{Black Holes in String
  Theory},''
\href{http://dx.doi.org/10.1016/0550-3213(89)90172-7}{{\em Nucl. Phys.}
  {\bfseries B311} (1989) 673--698}.

\bibitem{Nicolini:2005vd}
P.~Nicolini, A.~Smailagic, and E.~Spallucci, ``{Noncommutative geometry
  inspired Schwarzschild black hole},''
  \href{http://dx.doi.org/10.1016/j.physletb.2005.11.004}{{\em Phys. Lett.}
  {\bfseries B632} (2006) 547--551},
\href{http://arxiv.org/abs/gr-qc/0510112}{{\ttfamily arXiv:gr-qc/0510112
  [gr-qc]}}.

\end{thebibliography}\endgroup

\bibliographystyle{./utphys1}

\end{document}